\documentclass[12pt,preprint]{aastex6}

\begin{document}

\def\htwo{\rm H_2}
\def\hthree{\rm H_3^+}
\def\htwod{\rm H_2D^+}
\def\dtwoh{\rm D_2H^+}
\def\dthree{\rm D_3^+}
\def\diaz{\rm N_2H^+}
\def\ammo{\rm NH_3}
\def\Harpoons{\mathop{\rightleftharpoons}\limits}
\def\arrow{\mathop{\rightarrow}\limits}
\def\degr{\ensuremath{^\circ}}
\def\arcmin{\ensuremath{^\prime}}
\def\arcsec{\ensuremath{^{\prime\prime}}}
\def\kms{\rm km\,s^{-1}}

\title{Detection of interstellar ortho-$\dtwoh$ with SOFIA}

\author{Jorma Harju\altaffilmark{1,2,6}}

\author{Olli Sipil\"a\altaffilmark{1}}

\author{Sandra Br\"unken\altaffilmark{3}}

\author{Stephan Schlemmer\altaffilmark{3}}

\author{Paola Caselli\altaffilmark{1}}

\author{Mika Juvela\altaffilmark{2}}

\author{Karl M. Menten\altaffilmark{4}}

\author{J\"urgen Stutzki\altaffilmark{3}}

\author{Oskar Asvany\altaffilmark{3}}

\author{Tomasz Kami{\'n}ski\altaffilmark{4,5}}

\author{Yoko Okada\altaffilmark{3}}

\author{Ronan Higgins\altaffilmark{3}}

\altaffiltext{1}{Max-Planck-Institut f\"ur extraterrestrische Physik,
  Gie{\ss}enbachstra{\ss}e 1, 85748 Garching, Germany}

\altaffiltext{2}{Department of Physics, P.O. BOX 64, 00014 University
  of Helsinki, Finland}

\altaffiltext{3}{I. Physikalisches Institut, Universit\"at zu K\"oln,
  Z\"ulpicher Stra{\ss}e 77, 50937 K\"oln, Germany}

\altaffiltext{4}{Max-Planck-Institut f\"ur Radioastronomie,
  Auf dem H\"ugel 69, 53121 Bonn, Germany}

\altaffiltext{5}{Harvard-Smithsonian Center for Astrophysics, 60 Garden Street,
    Cambridge MA 02138, USA}

\altaffiltext{6}{harju@mpe.mpg.de}

\begin{abstract}

  We report on the detection of the ground-state rotational line of
  ortho-$\dtwoh$ at 1.477 THz ($203\,\mu$m) using the German REceiver
  for Astronomy at Terahertz frequencies (GREAT) onboard the
  Stratospheric Observatory For Infrared Astronomy (SOFIA).  The line
  is seen in absorption against far-infrared continuum from the
  protostellar binary IRAS 16293-2422 in Ophiuchus.  The para-$\dtwoh$
  line at 691.7 GHz was not detected with the APEX telescope toward
  this position. These $\dtwoh$ observations complement our previous
  detections of para-$\htwod$ and ortho-$\htwod$ using SOFIA and
  APEX. By modeling chemistry and radiative transfer in the dense core
  surrounding the protostars, we find that the ortho-$\dtwoh$ and
  para-$\htwod$ absorption features mainly originate in the cool ($T <
  18$ K) outer envelope of the core. In contrast, the ortho-$\htwod$
  emission from the core is significantly absorbed by the ambient
  molecular cloud. Analyses of the combined $\dtwoh$ and $\htwod$ data
  result in an age estimate of $\sim 5\times10^5$ yr for the core, with an
  uncertainty of $\sim 2\times10^5$ yr. The core material has
  probably been pre-processed for another $5\times10^5$ years in
  conditions corresponding to those in the ambient molecular cloud.
  The inferred time scale is more than ten times the age of the
  embedded protobinary. The $\dtwoh$ and $\htwod$ ions have large and
  nearly equal total (ortho+para) fractional abundances of $\sim
  10^{-9}$ in the outer envelope. This confirms the central role of
  $\hthree$ in the deuterium chemistry in cool, dense gas, and adds
  support to the prediction of chemistry models that also $\dthree$
  should be abundant in these conditions.

\end{abstract}


\section{Introduction} \label{sec:intro}

The deuterated variants of the $\hthree$ ion are believed to be the
most important agents of deuterium fractionation in cool ($T<20$ K)
interstellar clouds \citep{1989ApJ...340..906M}. Deuteration driven by
$\hthree$ should be particularly efficient in cold, dense cores.
Firstly, substitution of H by D is favored in cold gas because it
lowers the zero-point vibrational energy of a molecule.  Secondly,
$\hthree$ and its isotopologs are predicted to be enhanced greatly in
conditions where common molecules like CO and N$_2$ freeze 
onto dust grains (\citealt{1984ApJ...287L..47D};
\citealt{2003A&A...403L..37C}; \citealt{2003ApJ...591L..41R};
\citealt{2004A&A...418.1035W}; \citealt{2004A&A...424..905R}). These
chemistry models predict that gas with a large degree of CO depletion
can have abundances of $\htwod$ and $\dtwoh$ similar to that of
$\hthree$, and that $\dthree$ can become the most abundant molecular
ion.  The detection of multiply deuterated species in dense cores
supports strongly these ideas \citep{2005A&G....46b..29M}. On the
other hand, it has proven difficult to determine observationally the
total abundances of deuterated isotopologs of $\hthree$.

\setcounter{table}{0}
\begin{table*}[tbp] 
\renewcommand{\thetable}{\arabic{table}}
\centering
\caption{Ground-state rotational transitions of ortho and para
$\htwod$ and $\dtwoh$. Columns 4 and 5 give the upper state energies
and the Einstein coefficients for spontaneous emission.} 
\begin{tabular}{llllll}
\tablewidth{0pt}
\hline
\hline
Species & Transition & Frequency & $E_{\rm upper}$ & $A_{\rm ul}$ & 
Reference \\
        & ($J_{K_aK_c}^{\rm u} - J_{K_aK_c}^{\rm l}$) & (GHz) & (K) & (s$^{-1}$) & \\
\hline
para-$\htwod$ & $1_{01} - 0_{00}$ &  1370.1 &  65.8 & $4.04\times10^{-3}$ & a,c \\
ortho-$\htwod$ & $1_{10} - 1_{11}$ &  372.4 & 104.3 & $1.22\times10^{-4}$ & b,c \\
ortho-$\dtwoh$ & $1_{11} - 0_{00}$ & 1476.6 & 70.9 & $3.30\times10^{-3}$ &  a,c \\
para-$\dtwoh$ & $1_{10} - 1_{01}$  &  691.7 & 83.4 & $5.09\times10^{-4}$ &  b,c \\
\hline
\end{tabular}
\label{tab:trans}

a \cite{2008PhRvL.100w3004A}, b \cite{2005JMoSp.233....7A}, 
c \cite{2017JMoSp.332...33J}

\end{table*}

The asymmetric isotopologs $\htwod$ and $\dtwoh$ have permanent
electric dipole moments. Their ground-state rotational transitions lie
at submillimeter and far-infrared wavelengths. Like molecular
hydrogen, $\htwod$ and $\dtwoh$ exist as ortho and para modifications,
depending on the symmetries of the nuclear spin functions of the two
protons or deuterons. The lowest rotational transitions of the four
species are listed in Table~\ref{tab:trans}. While ortho-$\htwod$ has
been observed in several dense clouds (e.g.,
\citealt{2008A&A...492..703C}; \citealt{2012ApJ...751..135P}),
para-$\htwod$ has only been detected once so far
\citep{2014Natur.516..219B}.  The para modification of $\dtwoh$ has
been previously detected toward two starless cores
(\citealt{2004ApJ...606L.127V}; \citealt{2011A&A...526A..31P}),
whereas the previous search for ortho-$\dtwoh$ with Herschel/HIFI was
unsuccessful \citep{2012A&A...547A..33V}. In addition to astrochemical
interest, observations of $\htwod$ and $\dtwoh$ can also be used to
determine the time-scales of the earliest phases of
star-formation. This is made possible by the fact that both the
fractional abundances and the ortho/para ratios of these ions are
closely linked to o/p-$\htwo$, which is supposed to be 3 at the
formation on grains, but to decrease continually, albeit slowly, in
cold gas through proton exchange reactions
(\citealt{2004A&A...418.1035W}; \citealt{2011ApJ...739L..35P}).

In this paper we report on the observations of ortho- and
para-$\dtwoh$ rotational ground-state lines towards IRAS 16293-2422,
hereafter IRAS16293, a dense core in Ophiuchus hosting a protobinary
(\citealt{1989ApJ...337..858W}; \citealt{2016A&A...595A.117J}).
Besides providing the first detection of interestellar ortho-$\dtwoh$,
these observations complement our previous observations of ortho- and
para-$\htwod$ toward IRAS16293 \citep{2014Natur.516..219B}, which
allowed the determination of the o/p-$\htwod$ and the total
$\htwod$ abundance in the core's envelope. We used these quantities in
conjunction with chemical models to infer that the dense core has been
chemically processed for at least one million years. The inclusion of
ortho- and para-$\dtwoh$ to the observed species permits testing of
this estimate, and puts also our understanding of deuterium
chemistry to the proof. 

The paper is organized as follows. In Section 2, we describe the new
observations of the ortho- and para-lines of $\dtwoh$. The physical
and chemical model used for the interpretation of the observed spectra
is described in Section 3. The results of the modeling and the
comparison with the observations are discussed in detail in Section 4,
followed by a general discussion and conclusions in Section 5.

\section{Observations} \label{sec:obs}

\subsection{SOFIA} \label{subsec:sofia}

We observed the ortho-$\dtwoh$ ground-state transition $J_{K_aK_c} =
0_{00} \rightarrow 1_{11}$ at 1476.6 GHz ($\lambda=203\,\mu$m) toward
the protostellar binary embedded in IRAS16293 using the upgraded
far-infrared receiver upGREAT (\citealt{2012A&A...542L...1H};
\citealt{2016A&A...595A..34R})\footnote{The latest developments are
  described at www3.mpifr-bonn.mpg.de/div/submmtech} onboard SOFIA
\citep{2012ApJ...749L..17Y}. The measurement was made on 2015 May 22,
during one of the commissioning flights of upGREAT, which started from
Palmdale, California. The target position was R.A. $16^{\rm h}32^{\rm
  m}22\fs9$, Dec. $-24\degr28\arcmin39\arcsec$ (J2000), lying
  $3\arcsec$ south of the component A of the protobinary. The same
  position was observed in para-$\htwod$ by
  \cite{2014Natur.516..219B}. With this choice, based on the Herschel
  $250\,\mu$m map, we wanted to maximise the continuum signal. We
used the GREAT L1b channel (1.42-1.52 THz) with the FFTS4G backend,
which has a bandwidth of 4 GHz and a channel spacing of 283 kHz (58
m\,s$^{-1}$). For the sky subtraction, the instrument was operated in
the double beam-switch mode. The chopping frequency was 1 Hz, and the
chop amplitude was $20\arcsec$ which provided a total beam throw of
$40\arcsec$ in a horizontal direction.  The on-source integration time
was 25 min. The average system temperature was $T_{\rm sys}= 4080$ K.
The forward and main beam efficiencies for the L1 channel of GREAT are
0.97 and 0.64, respectively. The half-power beam width (HPBW) of the
2.5-m SOFIA telescope is $18.1\arcsec$ at 1476.6 GHz.

\begin{figure}
\figurenum{1}
\plotone{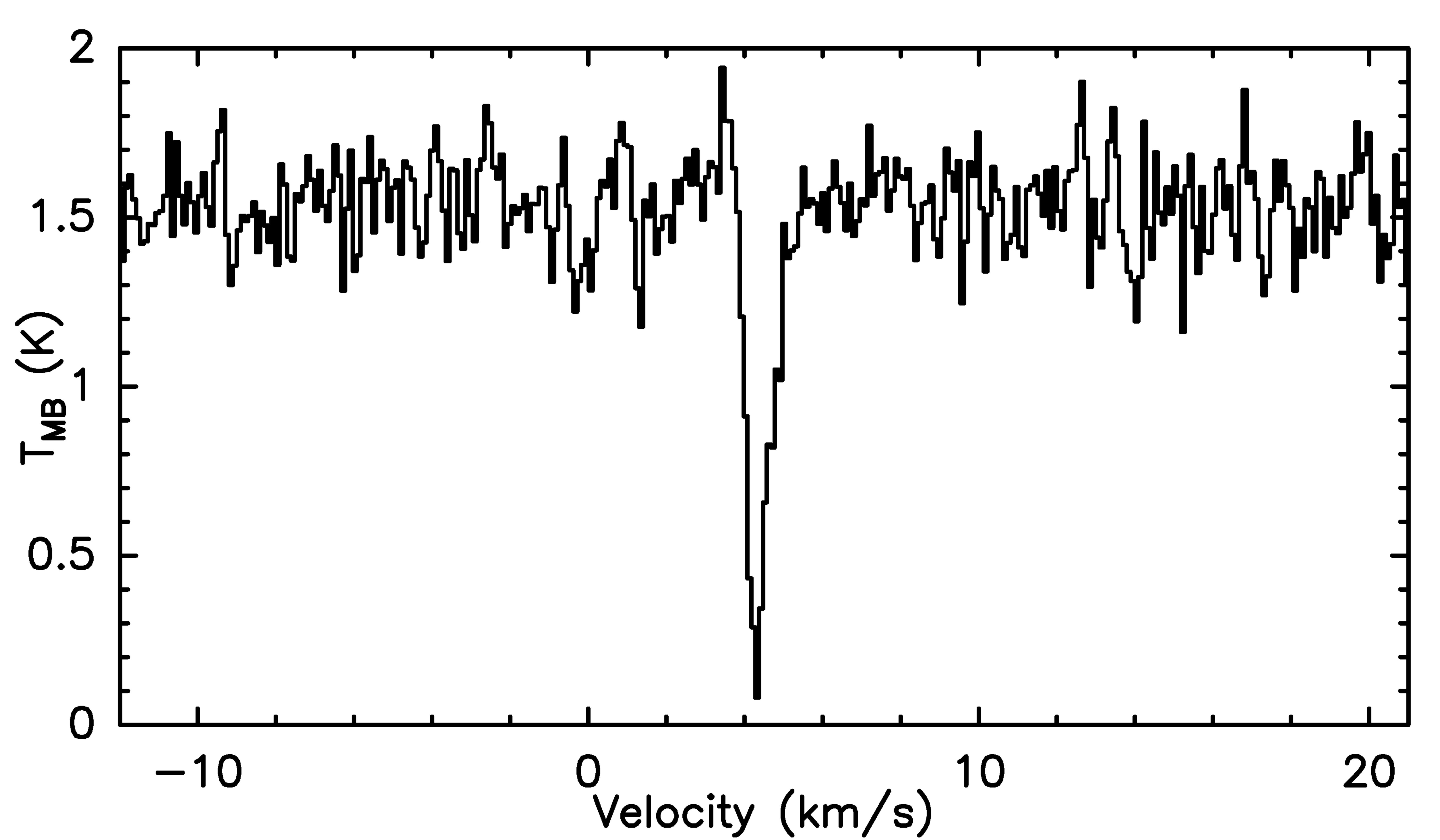}
\caption{Ground-state rotational line of ortho-$\dtwoh$ at 1476.6 GHz 
observed with SOFIA/GREAT toward IRAS16293.}
  \label{fig:sofia_od2h+}
\end{figure}

The GREAT receiver admits two sidebands, while the absorption
  feature appears in only one of them.  We calibrated the absorption
  spectrum and the continuum separately, and combined these afterwards
  so that the correct intensity scale is recovered.  In the continuum
  calibration, the contribution from both sidebands was taken into account.
  The derived continuum level is $T_{\rm mb,c} = 1.54$ K on the
  Rayleigh-Jeans temperature scale, which corresponds to 960
  Jy\,beam$^{-1}$.  This constant continuum level was added to the
  absorption spectra to get a final spectrum.  The calibrated
ortho-$\dtwoh$ spectrum is shown in Figure~\ref{fig:sofia_od2h+}. The
continuum from the protostellar core is almost completely absorbed in
the line center.

\subsection{APEX} \label{subsec:apex}

The para-$\dtwoh$ ground-state line at 691.7 GHz
($J_{K_aK_c}=1_{10}\rightarrow 1_{01}$) was observed using the CHAMP+
array receiver \citep{2006SPIE.6275E..0NK} on Atacama Pathfinder
Experiment 12 meter diameter submillimeter telescope (APEX)
\citep{2006A&A...454L..13G}. The receiver has seven pixels in the
waveband covering the target line. The footprint consists of a central
pixel surrounded by a hexagonal pattern. The pixels are separated by
about $20\arcsec$, with the central pixel pointed at the
  protostar IRAS 16293A, R.A. $16^{\rm h}32^{\rm m}22\fs9$,
Dec. $-24\degr28\arcmin36\arcsec$ (J2000).  The center position
  lies $3\arcsec$ off from the position observed with SOFIA. This
  separation caused by a mistake is small compared to the SOFIA beam.
The beamwidth of APEX is $9\arcsec$ at 691.7 GHz, and the main beam
efficiency is 0.47. The observations were made on 2013 August 4, in
excellent conditions (PWV 0.35 mm). The observing mode was wobbler
switching with a beam throw of $150\arcsec$. The total on-source
integration time was 40 min at a system temperature of 1170 K. The
CHAMP+ Array FFTS with a velocity resolution of 0.3175 km\,s$^{-1}$
was used as the backend.  We reached an rms noise of 0.10~K on the
$T_{\rm mb}$ scale in the central pixel.  No line was detected in any
of the pixels. The continuum level was measured to be $T_{\rm mb,c} =
2.5\pm0.1$ K, which corresponds to $66\pm3$ Jy\,beam$^{-1}$.

\section{Modeling}

\subsection{Physical model} \label{subsec:physical} 

\begin{figure*}
\figurenum{2}
\unitlength=1mm
\begin{picture}(160,70)
\put(0,0){
\begin{picture}(0,0) 
\includegraphics[width=9cm]{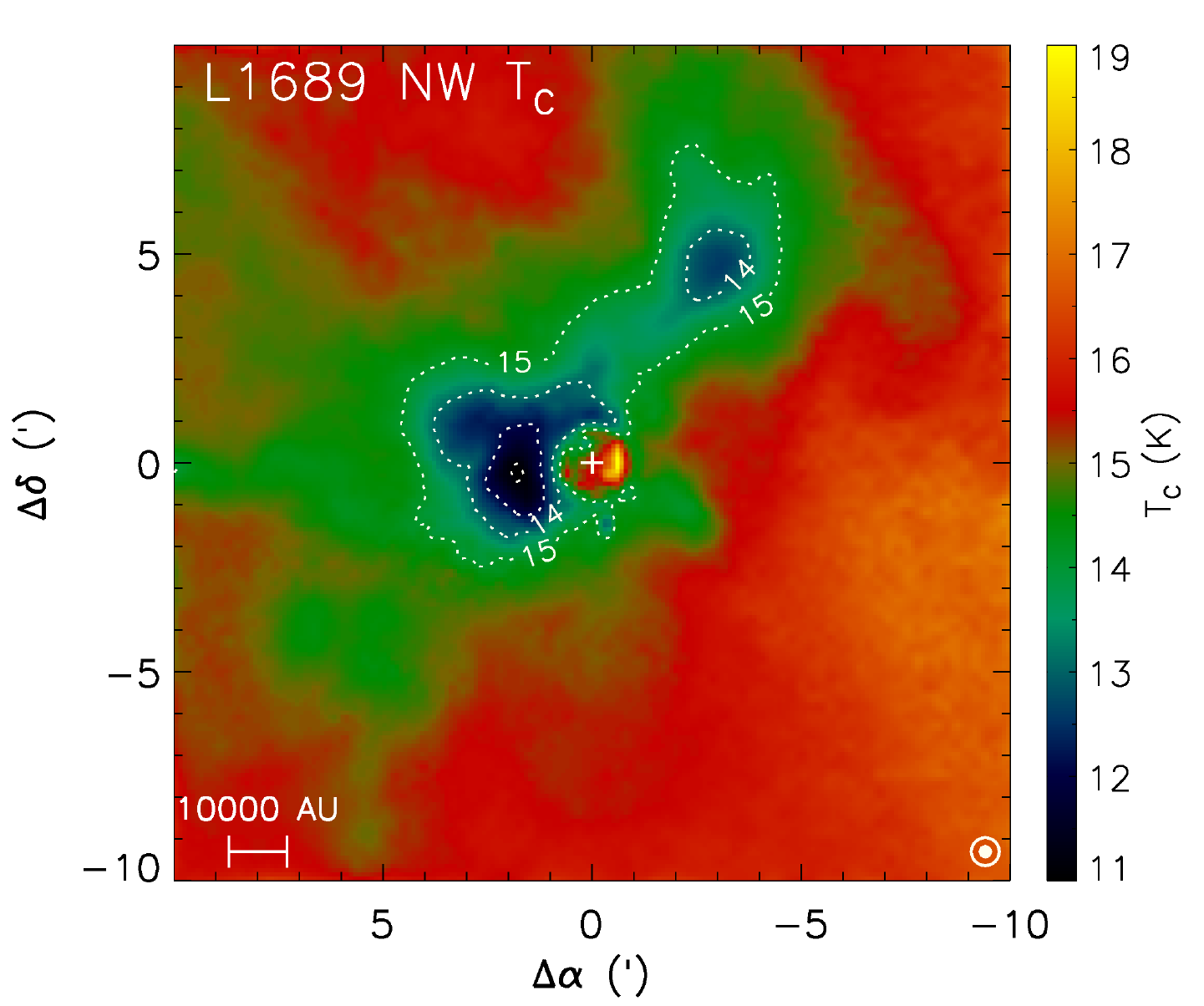}
\end{picture}}
\put(90,0){
\begin{picture}(0,0) 
\includegraphics[width=9cm]{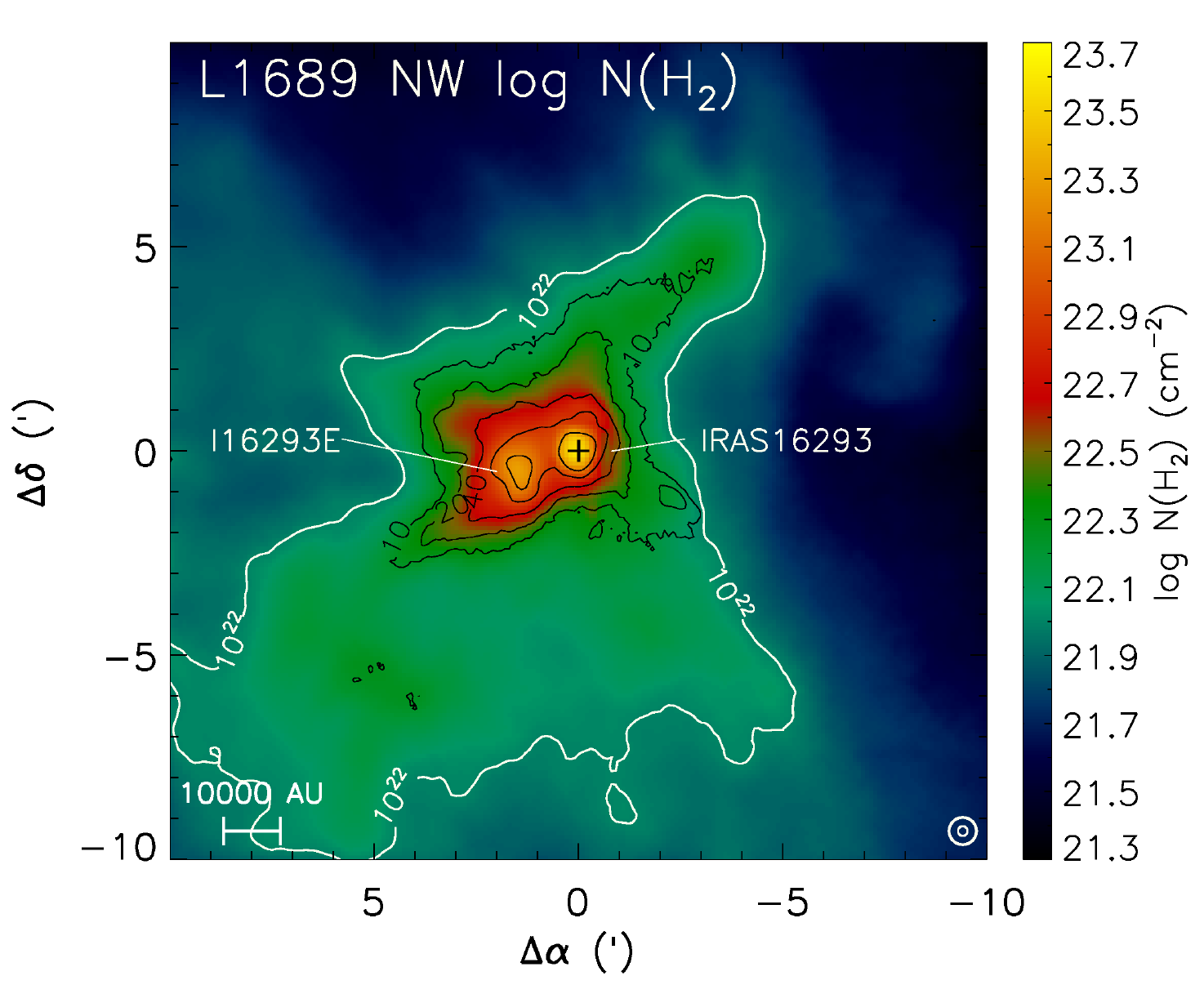}
\end{picture}}
\put(71,66){\makebox(0,0){\large \bf \color{white} a)}}
\put(161,66){\makebox(0,0){\large \bf \color{white} b)}}
\end{picture}

\caption{Dust colour temperature,  $T_{\rm C}$ (a), and the
  $\htwo$ column density, $N(\htwo)$ (b), maps of the surroundings
  of IRAS16293 as derived from Herschel/SPIRE maps at 250, 350, and
  500 $\mu$m. The plus sign indicates the position where the present 
  observations were pointed at. The filled circle in the bottom right 
  corresponds to the SOFIA beam ($\sim 18\arcsec$). The larger circle 
  represents the resolution of the $N(\htwo)$ and $T_{\rm C}$ maps ($\sim
  40\arcsec$). The distribution of the 850 $\mu$m emission observed with 
  SCUBA-2 is indicated with black contours on the $N(\htwo)$ map.  The contour
  levels are 10, 20, 40, 80, and 160 MJy\,sr$^{-1}$. The $14^{\prime\prime}$
  resolution of the SCUBA-2 $850\,\mu$m map is indicated with the
  smaller open circle. The bright linear structure seen near the center of
  the $T_{\rm C}$ map is an artifact resulting from the interpolation
  of the surface brightness maps to a common grid. It is caused by a
  steep surface brightness gradient on the eastern side of IRAS16293.}
\label{fig:l1689nw}
\end{figure*}

The protostellar core IRAS16293 has been studied extensively. The
object lies in the northwestern part of the dark cloud Lynds 1689, which
is part of the $\rho$ Ophiuchi molecular cloud complex. The distance
to the $\rho$ Oph cloud is approximately 120 pc
(\citealt{2008ApJ...675L..29L}; \citealt{2008A&A...480..785L}).  The
overall structure of the cloud surrounding IRAS16293 and its
companion, the starless core ``I16293E'' \citep{2004ApJ...608..341S},
is illustrated in Figure~\ref{fig:l1689nw}, which shows the dust colour
temperature ($T_{\rm C}$) and the $\htwo$ column density ($N(\htwo)$)
maps of a $20\arcmin \times 20\arcmin$ region centered on
IRAS16293. These maps cover the high-column density part of the region
called L1689-NorthWest by \cite{2006MNRAS.368.1833N}. The $T_{\rm C}$
and $N(\htwo)$ maps are derived from far-infrared images observed by
{\sl Herschel} \citep{2010A&A...518L...1P}. Contours of the 850 $\mu$m
emission map from the SCUBA-2 survey of \cite{2015MNRAS.450.1094P} are
superposed on the $N(\htwo)$ map.

The Herschel/SPIRE images were extracted from the pipeline-reduced
images of the Ophiuchus complex made in the course of the Herschel
Gould Belt Survey \citep{2010A&A...518L.102A}. The data are downloaded
from the Herschel Science Archive
(HSA)\footnote{\url{www.cosmos.esa.int/web/herschel/science-archive}}. We
calculated the $T_{\rm C}$ and $N(\htwo)$ distributions using only the
three SPIRE \citep{2010A&A...518L...3G} bands at $250\,\mu$m,
$350\,\mu$m, and $500\,\mu$m, for which the pipeline reduction includes
zero-level corrections based on comparison with the Planck satellite
data. A modified blackbody function with the dust emissivity spectral
index $\beta=2$ was fitted to each pixel after smoothing the
$250\,\mu$m and $350\,\mu$m images to the resolution of the
$500\,\mu$m image ($\sim 40\arcsec$), and resampling all images to the
same grid. For the dust emissivity coefficient per unit mass of gas,
we adopted the value from \cite{1983QJRAS..24..267H}, $\kappa_{250\mu
  \rm m}=0.1$ cm$^{2}$g$^{-1}$.

The core IRAS16293 is visible as a roundish peak on the column density
and the 850 $\mu$m surface brightness maps shown in
Figure~\ref{fig:l1689nw}. The general consensus on the structure of
IRAS16293 is that it harbors a double Class 0 protostar, IRAS16293A
and IRAS16293B with a separation of about 600 AU, surrounded by a
massive envelope ($\sim 2\,M_\odot$) with steeply decreasing
temperature and density distributions (e.g.,
\citealt{1989ApJ...337..858W}; \citealt{1990ApJ...365..269L};
\citealt{2004ApJ...608..341S}; \citealt{2005ApJ...631L..77J};
\citealt{2010A&A...519A..65C}; \citealt{2016A&A...595A.117J}).

\begin{figure}[htbp]
\figurenum{3}
\plotone{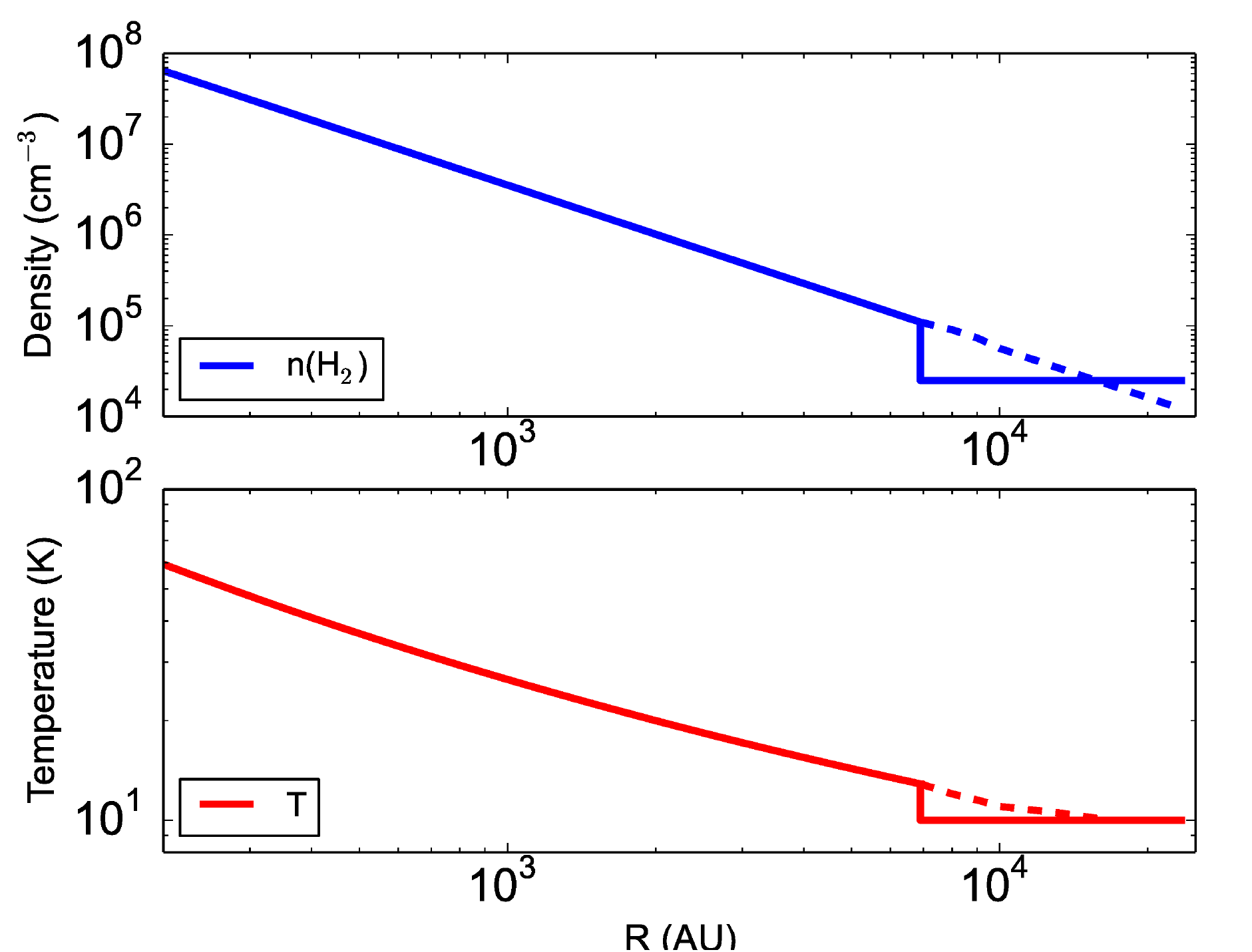}
\caption{Assumed density and temperature distributions of
  IRAS16293.  The core model (up to $R=6900$ AU) is adopted from
  \citet{2010A&A...519A..65C}.  The core is likely to be embedded in 
  the L1689N molecular cloud. The two alternative ambient cloud models 
  used here are 1) a homogeneous dark cloud-type model (solid lines), and 
  2) extension of the density power-law and the corresponding temperature
  profile from Crimier et al.  (dashed lines).}
\label{fig:crimier_ext_model}
\end{figure}

We adopt the frequently used physical model of IRAS16293, derived by
\citet{2010A&A...519A..65C}, where the radial density distribution of
the protostellar envelope is described by a power law, $n(\htwo)
\propto r^{-1.8}$, between the central cavity (27 AU) and the outer
edge (6900 AU), corresponding to $\sim 60\arcsec$ at the distance 120
pc. The model is not accurate in the center of the core, but describes
well the structure seen with a single-dish telescope. 
The heating of the inner envelope is dominated by the central
protostars for which Crimier et al. derived a bolometric
luminosity of $22\, L_\odot$. The density and the kinetic temperature
distributions of the model are plotted in
Figure~\ref{fig:crimier_ext_model}.  The gas kinetic temperature and the
dust temperature are assumed to be equal in the core envelope. 

Two models for the ambient molecular cloud were tested. Firstly, we
assumed that the core is surrounded by a homogeneous cloud with
physical conditions considered to be typical for a dark cloud.  In
\cite{2014Natur.516..219B}, the temperature and density of the ambient
cloud were fixed to $n(\htwo)=10^4$ cm$^{-3}$ and $T=10$ K. Here we
start with these values, but also explore the ranges $T=
10-14$ K and $n(\htwo) = 1-4 \times 10^4$ cm$^{-3}$.  In the second
ambient cloud model, the Crimier et al. density and temperature
distributions are extrapolated to larger distances from the center of
the core. The outer boundary of the ambient cloud model is determined
by the assumption that this component causes a visual extinction of
$A_{\rm V} =10^{\rm mag}$ to the outer edge of the dense core. The
cloud constructed in this way extends to a distance of approximately
24000 AU ($200\arcsec$) from the core center. For the homogeneous dark
cloud model, the distances range from 15000 AU ($120\arcsec$) to 38000
AU ($320\arcsec$), depending on the density. The chosen value for the
external extinction, $A_{\rm V} =10^{\rm mag}$, is arbitrary but
possible, as the $\htwo$ column density around the core is of the
order of $10^{22}$ cm$^{-2}$.

The near side of the ambient cloud can be identified with the
``foreground cloud'' invoked in recent Herschel/HIFI absorption line
studies (\citealt{2012A&A...539A.132C}; \citealt{2013A&A...553A..75C};
\citealt{2014MNRAS.441.1964B}). In accordance with these studies, we
assume in the spectral line modeling that the foreground component can
have a different LSR velocity than the dense core, which gives rise to
asymmetric absorption profiles as observed in our SOFIA spectra of 
ortho-$\dtwoh$ and para-$\htwod$.

\subsection{Chemistry model} \label{subsec:chemistry}

The chemical abundances in different parts of IRAS16293 and the
ambient cloud were calculated using a chemistry model containing
reaction sets for both gas-phase and grain surface chemistry. The code
is described in detail in \citet{2015A&A...578A..55S} and in
\citet{2015A&A...581A.122S}.  The chemistry model distinguishes
between the different nuclear spin states of light hydrogen molecules,
nitrogen hydrides, water, and their deuterated forms. The reaction
rate coefficients for the $\hthree + \htwo$ isotopic system, which is
most relevant for the present study, are adopted from
\citet{2009JChPh.130p4302H}.  For dissociative recombination reactions
of $\hthree$ and its deuterated forms with electrons we used the rate
coefficients presented in \citet{2009A&A...494..623P}.  In the present
work we use the KIDA gas-phase network \citep{2015ApJS..217...20W}
as the basis upon which the deuterium and spin-state chemistry
is added with the prescriptions detailed in \cite{2015A&A...578A..55S} 
and \cite{2015A&A...581A.122S}.

The method of calculation is similar to that described in
\citet{2014Natur.516..219B}. The physical model is divided into
concentric shells where the density and temperature are assumed to be
constant. Chemical evolution is calculated separately in each
  shell. The model is pseudo-time dependent, i.e., we follow
the chemical evolution assuming that the protostellar envelope and the
ambient cloud are static.  

We use different assumptions about initial chemical composition of the
cloud. In one of the models, we assume that the gas is initially
atomic, with the exceptions of hydrogen and deuterium, which are
completely bound to $\htwo$ and HD. We also calculated a series of
models, for which the initial composition depends on the processing time
in conditions corresponding to the outskirts of the ambient cloud
($n(\htwo)\sim 10^4$ cm$^{-3}$, $T\sim 14$ K). For the elemental
abundances, we have adopted the set of low-metal elemental abundances
labelled EA1 in \cite{2008ApJ...680..371W}. We have fixed the
cosmic-ray ionization rate to $\zeta = 1.3\times10^{-17}$ s$^{-1}$, the
grain radius to $a_{\rm g} = 0.1\,\mu$m, and the initial
ortho/para-$\htwo$ to $10^{-3}$, corresponding to a spin temperature
of $T_{\rm spin} \sim 18$~K.

\subsection{Simulation of spectral lines} \label{subsec:radiative}

The abundance distributions extracted from the chemistry model are
used, together with the density, temperature, and velocity profiles of
the core, as input for predicting observable spectral line
profiles. Our radiative transfer calculations include the emission and
absorption by dust, and the effect of dust emission on the line
excitation.  We use programs described in \citet{1997A&A...322..943J}
and \citet{2005A&A...440..531J}. The input parameters of the physical
model are the cloud density structure, the luminosity of the central
protostar, the intensity of the interstellar radiation field (ISRF),
and the dust opacity spectrum. We adopted the dust opacity model of
\citet{1994A&A...291..943O} for dust grains with thin ice coatings
processed at the density $n(\htwo)=10^5\,{\rm cm}^{-3}$.  The spectrum
of the unattenuated interstellar radiation field (ISRF) was taken from
\citet{1994ASPC...58..355B}.  The program produces a high angular
resolution spectral map of the source model, including the dust
continuum. For comparison with observations, the map is smoothed to a
given resolution. The excitation of the rotational transitions of
$\dtwoh$ and $\htwod$ in collisions with para- and ortho-$\htwo$ are
calculated using the state-to-state rate coefficients from
\citet{2009JChPh.130p4302H}. The hyperfine patterns of the lines are
adopted from \cite{1997MolPh..91..319J}. These are, however,
overwhelmed by the Doppler broadening.

\section{Results} \label{sec:results}

\subsection{Abundance distributions} \label{subsec:abundances}

\begin{figure}[htbp]
\figurenum{4}
\plotone{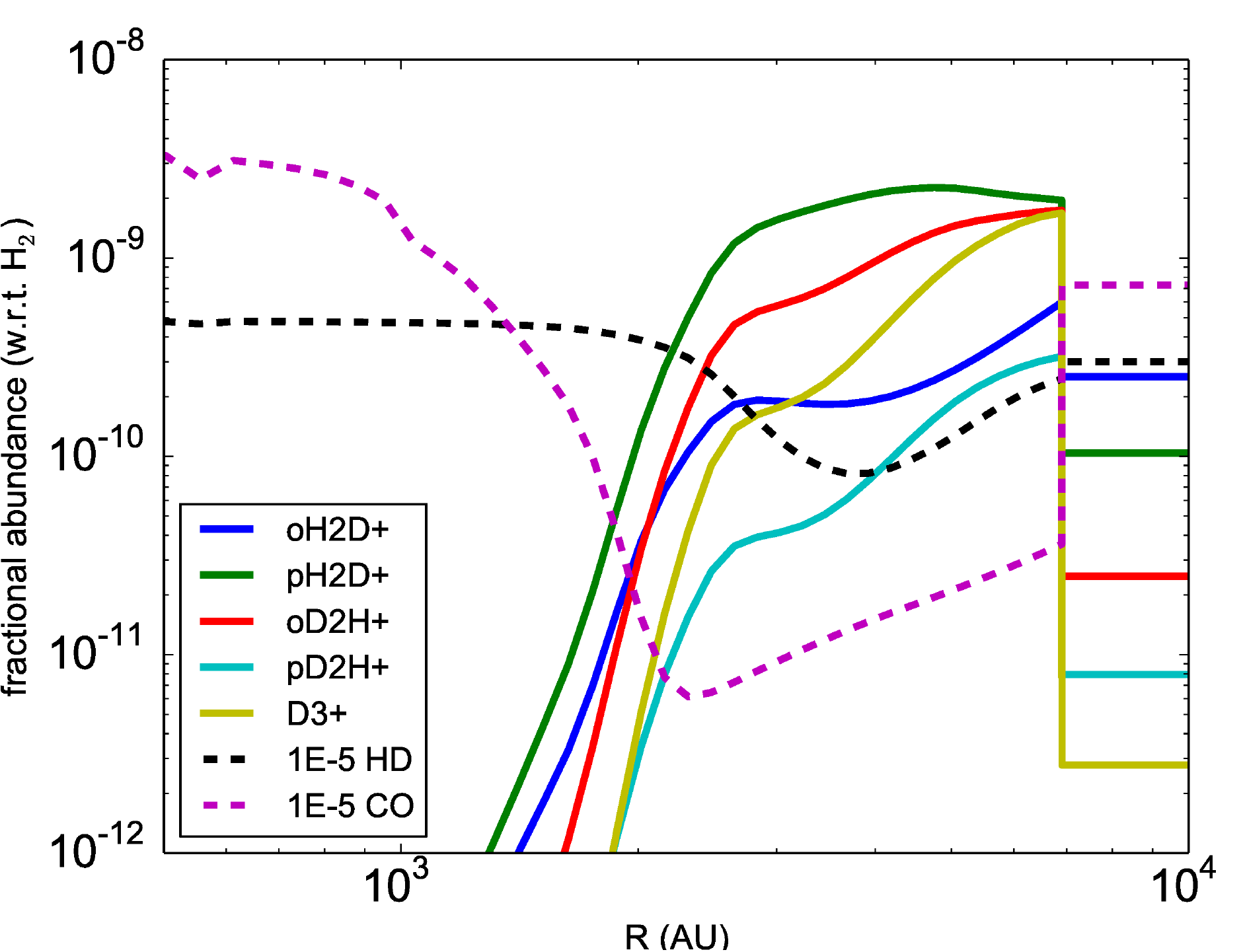}
\caption{Distributions of the $\htwod$, $\dtwoh$, and $\dthree$
  abundances (relative to $\htwo$) in the outer envelope of the core
  at the time $5\times10^5$ yr from the beginning of the simulation. Also
  shown are the fractional CO and HD abundances multiplied by
  $10^{-5}$.  The ambient cloud is assumed to be homogeneous with
  $n(\htwo)=10^4$ cm$^{-3}$, $T=10$ K.}
\label{fig:abu_rad_atomini}
\end{figure}

The abundances of deuterated ions can become significant only in cool,
shielded parts of the core envelope and the ambient dark cloud.  This
is illustrated in Figure~\ref{fig:abu_rad_atomini}, which shows the
radial abundance distributions of the deuterated forms of $\hthree$ at
an advanced stage ($t=5\times10^5$ yr) of the simulation that has been
started from an atomic initial composition (except for $\htwo$ and
HD). Also shown are the fractional CO and HD abundances.  Here we have
used the same ambient cloud model as \cite{2014Natur.516..219B} with
$n(\htwo)=10^4$ cm$^{-3}$ and $T=10$ K. The highest fractional
abundances of $\htwod$ and $\dtwoh$ ($X >10^{-10}$) are found in the
outer layers of the core envelope at radii $R \gtrsim 2500$ AU.  The
inner boundary marks the distance where the temperature decreases
below 20 K, making CO desorption inefficent. There is an abrupt change
in the abundances of $\htwod$, $\dtwoh$, and $\dthree$ at the
outer edge of the core because of the density drop; the chemical
evolution is slower in the ambient cloud. The abundances of deuterated
ions fall steeply near the extreme outer boundary of the ambient cloud
($A_{\rm V} \lesssim 5$, not shown).  This decrease is caused by
active CO production and a high ortho-$\htwo$ abundance close to the
surface of the cloud. Ultimately, CO is photodissociated in the
outermost layers with $A_{\rm V} < 1^{\rm mag}$.

\begin{figure*}[]
\figurenum{5}
\unitlength=1mm
\begin{picture}(160,68)
\put(0,0){
\begin{picture}(0,0) 
\includegraphics[width=9cm]{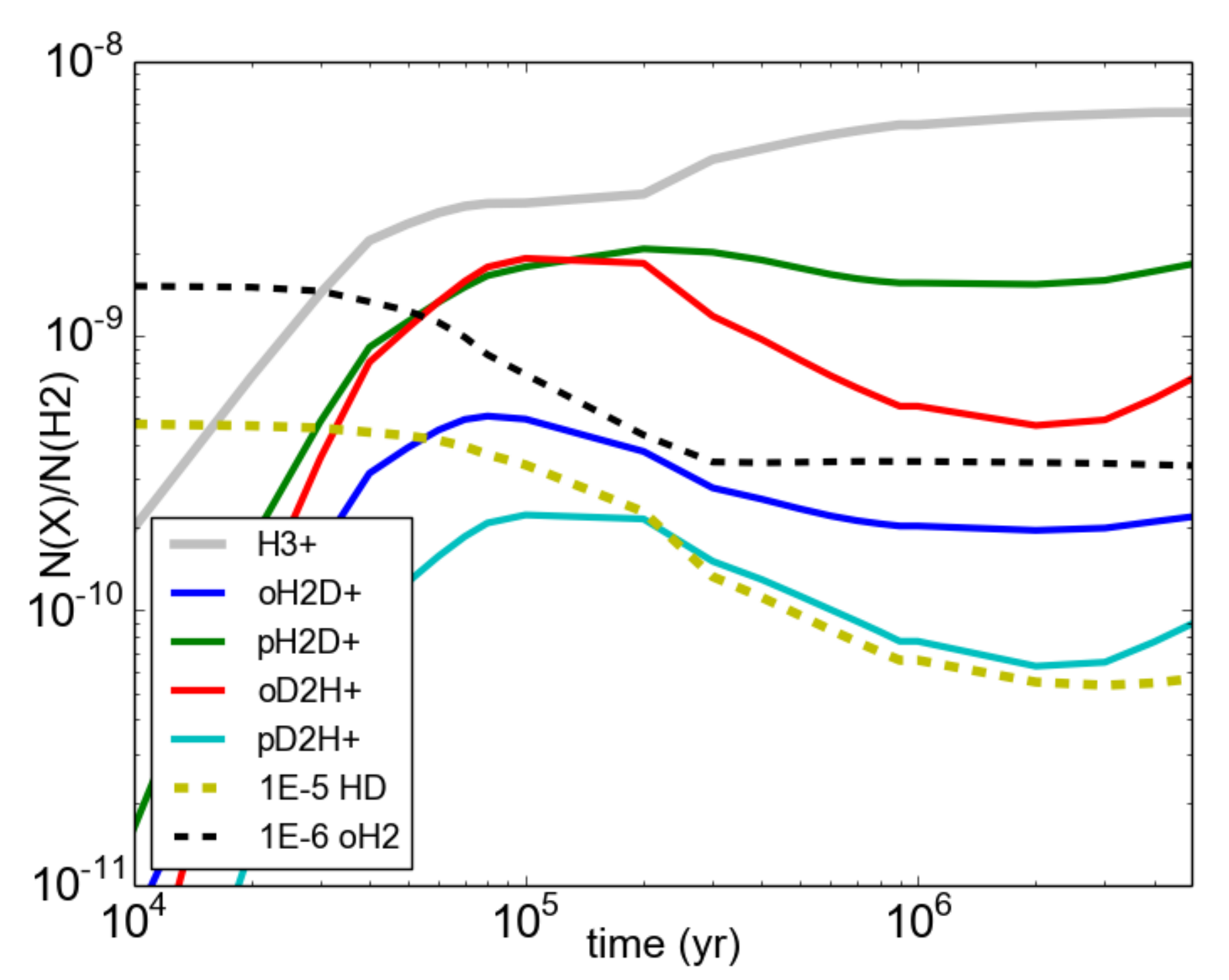}
\end{picture}}
\put(90,0){
\begin{picture}(0,0) 
\includegraphics[width=9cm]{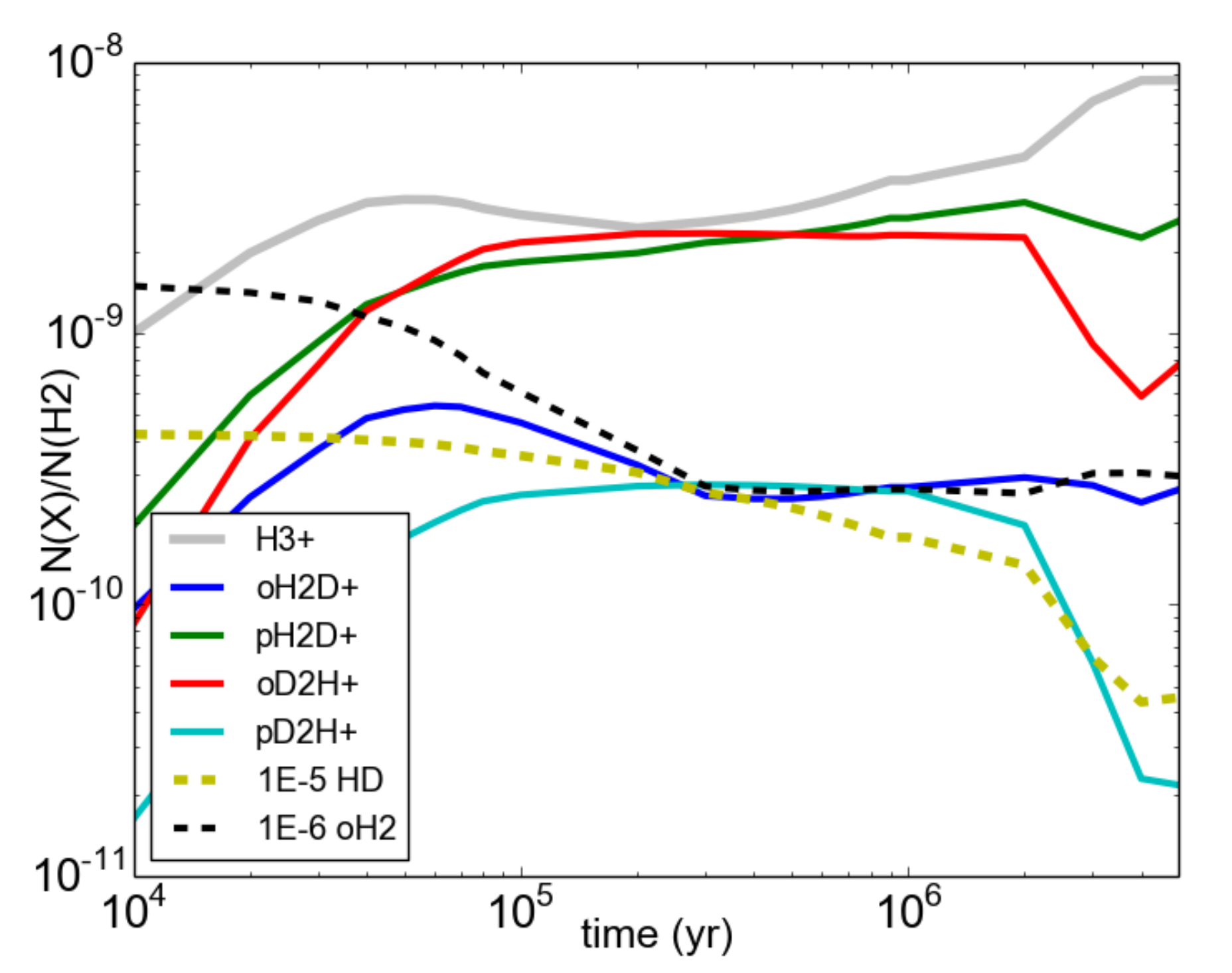}
\end{picture}}
\put(18,61){\makebox(0,0){\large \bf a)}}
\put(108,61){\makebox(0,0){\large \bf b)}}
\end{picture}
        \caption{Time evolution of the average fractional abundances
          of ortho- and para-$\dtwoh$ and ortho- and para-$\htwod$ in
          the outer layers ($R=2500-6900$ AU) of the protostellar
          envelope model using two different initial compositions: (a)
          atomic abundances and (b) abundances from the dark cloud
          stage after $5\times10^5$ yr. Also shown are the fractional abundances of
          $\hthree$, HD, and ortho-$\htwo$.}
\label{fig:abu_time}
\end{figure*}
  
We next examine the chemical evolution of the coolest part ($R\sim
2500-6900$ AU, $T<18$ K) of the envelope. The average abundances of ortho- and
para-$\htwod$ and $\dtwoh$ as functions of time using two different
initial gas compositions are shown in Figure~\ref{fig:abu_time}. Also
shown are the abundances of $\hthree$, HD, and ortho-$\htwo$. The fractional
abundances are column density-weighted averages, $N(X)/N(\htwo)$, over
the mentioned layers of the envelope.

The initial composition influences the evolution of the ortho-$\dtwoh$
abundance. When starting the simulation from unprocessed abundances
(we call this simulation ``a''), the ortho-$\dtwoh$ abundance
decreases substantially after an early peak at $t\sim 2\times10^5$ yr (see
Figure~\ref{fig:abu_time} a). On the other hand, if the chemistry is
first allowed to evolve for $5\times10^5$ yr (Figure~\ref{fig:abu_time} b; we
call this simulation ``b''), or any time of that order in dark cloud
conditions, the ortho-$\dtwoh$ abundance remains high until the late
stages of chemical evolution. The difference is related to a slower
depletion of HD and a faster decrease of o/p-$\htwo$ in simulation b
as compared with simulation a. This in turn can be attributed to a
higher $\hthree$ abundance in the beginning of simulation b.  In the
presence of HD, the $\hthree$ ion is efficiently deuterated, and all
its isotopologs contribute strongly to the ortho-para conversion of
$\htwo$.  The decrease of o/p-$\htwo$ further enhances deuterium
fractionation. Individual deuterium atoms are not deposited in
$\htwod$, $\dtwoh$ or $\dthree$ molecules for long, but they are
released by dissociative recombinations with electrons. The abundance
of D atoms in the gas phase and on grains is, therefore, higher in
simulation b than in simulation a. This is reflected in the
replenishment of HD from grains. In simulation b, the rate of HD
formation on grains is higher than in simulation a until very late
stages. To summarize, our interpretation is that the distribution of
deuterium between many different gas- and solid-phase species in the
beginning of simulation b helps to sustain high HD and $\dtwoh$
abundances longer than in simulation a, where all deuterium is
initially bound to HD. 

\begin{figure}[]
\figurenum{6}
\plotone{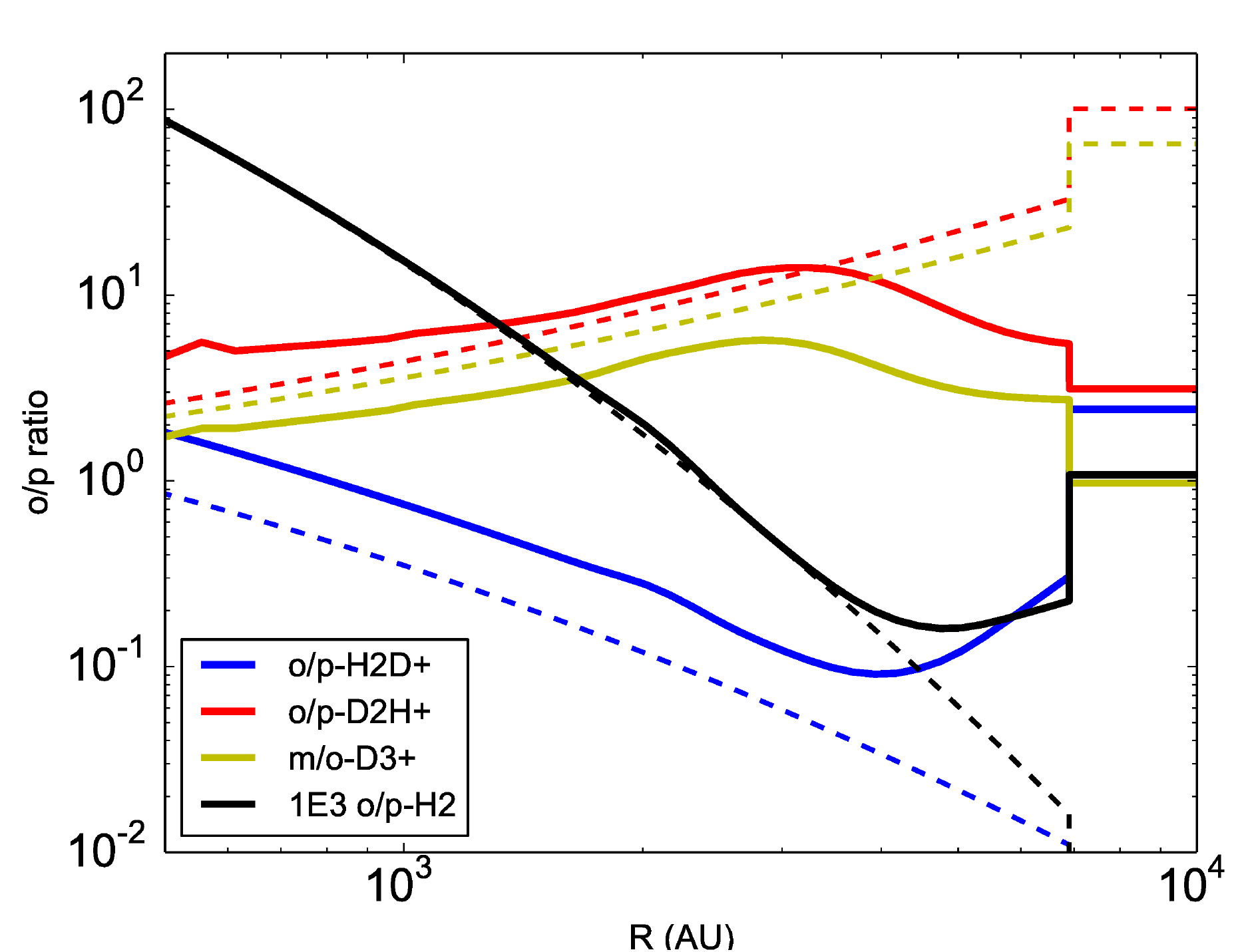}
\caption{Ortho/para ratios of $\dtwoh$, $\htwod$, and $\htwo$, and
  the meta/ortho ratio of $\dthree$ (solid lines) as functions of the
  radial distance from the core center. The corresponding ratios in
  thermal equilibrium are shown with dashed curves. The plot shows the
  situation at $t=5\times10^5$ yr in the simulation starting from a mainly
  atomic gas composition (simulation a). The ambient cloud is
  homogeneous at $n(\htwo)=10^4$ cm$^{-3}$, $T=10$ K.  Note that
  ortho-$\dtwoh$ and meta-$\dthree$ are the lowest-energy spin
  modifications of $\dtwoh$ and $\dthree$, whereas for $\htwo$ and
  $\htwod$, the para modifications represent the lowest energies.}
\label{fig:op_rad}
\end{figure}

The radial distributions of the o/p ratios of $\dtwoh$, $\htwod$, and
$\htwo$, and the meta/ortho (m/o) ratio of $\dthree$\footnote{The
  $\dthree$ ion has three permutation symmetry species in $S_3$: $E$
  (``ortho'', with nuclear spins $I=1$ or $I=2$), $A_1$
  (``meta'', $I=1,3$), and $A_2$ (``para'', $I=0$). The statistical
  weights (the numbers of possible nuclear spin states belonging to
  these species) are 16, 10, and 1 for ortho, meta, and para,
  respectively.}  are shown in Figure~\ref{fig:op_rad}. The ratios
correspond to the time $5\times10^5$ yr in the simulation presented in
Figure~\ref{fig:abu_time} (a), where the initial gas composition is
mainly atomic.  The dashed curves show the spin ratios in thermal
equilibrium. These are calculated from the partition functions, for
example o/p-$\htwo(T) = Q_{{\rm o}\htwo}(T)/Q_{{\rm p}\htwo}(T)$. At
low temperatures, the equilibrium ratios can be approximated by
o/p-$\htwo \approx 9\,\exp(-170.5/T)$, o/p-$\htwod \approx
9\,\exp(-86.4/T)$, o/p-$\dtwoh \approx 6/9\,\exp(+50.2/T)$, and
m/o-$\dthree \approx 10/16\,\exp(+46.5/T)$
(\citealt{2004A&A...427..887F}; \citealt{2009JChPh.130p4302H}).  The
modeled ortho/para ratios experience both radial and temporal changes.
In the dense and warm inner parts, o/p-$\htwo$ is thermalized, and
also o/p-$\htwod$, o/p-$\dtwoh$, and m/o-$\dthree$ follow their
thermal ratios (the perceptible offset of o/p-$\htwod$ from the
thermal ratio is discussed in Section~\ref{ss:opconv}). In the cool,
outer envelope the spin ratios deviate clearly from their thermal
values.  In the outermost layers with $T \leq 16$ K ($R \gtrsim 3500$
AU), all the plotted spin ratios are suprathermal, i.e., the species
with the higher ground-state energies, ortho-$\htwo$, ortho-$\htwod$,
para-$\dtwoh$, and ortho-$\dthree$, have clearly larger abundances
than expected from thermal equilibrium. The deviations in o/p-$\htwo$
and o/p-$\htwod$ from thermal ratios become smaller as time advances.
For o/p-$\dtwoh$ and m/o-$\dthree$, no substantial changes occur after
$10^5$ yr in this model. The spin ratio plot for simulation b (not
shown) is very similar to that shown in Figure~\ref{fig:op_rad} except
that the o/p ratios of $\htwo$ and $\htwod$ follow the thermal ratios
up to larger distances from the core center than in simulation a.

\begin{figure}[]
\figurenum{7}
\plotone{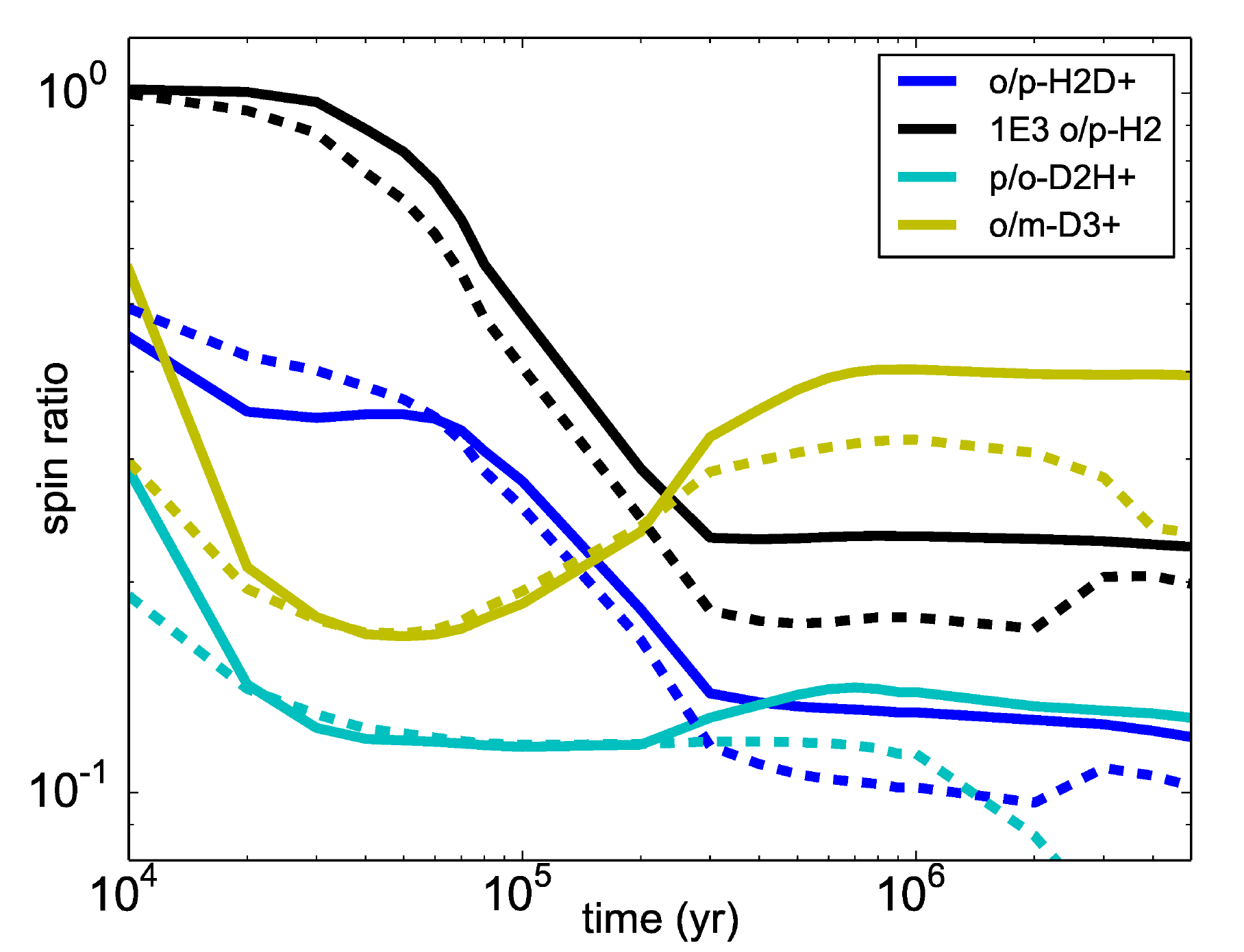} 
\caption{Average ortho/para ratios of $\htwod$ and $\htwo$,
  para/ortho-$\dtwoh$, and ortho/meta-$\dthree$ as functions of time in
  the outer envope of IRAS16293. The averages are calculated over the
  distance range $R=2500-6900$ AU from the center. Predictions from
  simulation a are shown with solid lines, and those from simulation b
  are shown with dashed lines.}
\label{fig:op_time}
\end{figure}

The time evolution of the averaged ortho/para ratios of $\htwo$,
$\htwod$, $\dtwoh$, and the ortho/meta (o/m) ratio of $\dthree$ are
presented in Figure~\ref{fig:op_time}. The averages are column density
ratios calculated over the outer envelope, $R = 2500-6900$ AU. The
figure shows the abundance ratios of the higher energy modification to
the lower energy modification for all molecules. The predicted
  ratios from simulation a, starting from atomic abundances, are shown
  with solid lines, and the ratios from simulation b, which starts
  from re-processed abundances, are shown with dashed lines. In
  simulation b, the ortho/para ratios of $\htwo$ and $\htwod$ decrease
  more steeply and reach lower values than in simulation
  a. Correspondingly, p/o-$\dtwoh$ and o/m-$\dthree$ in simulation b
  are lower at late times as compared to simulation a. In both
  simulations, o/p-$\htwod$ correlates closely with o/p-$\htwo$. As
discussed in \citet{2014Natur.516..219B}, this correlation is
established by proton exchange reactions between (ortho or
para)$\htwod$ and (ortho or para)$\htwo$. The o/p-$\htwod$ ratio as a
function of o/p-$\htwo$ can be calculated to a great accuracy from the
analytical formula given in \citet{2014Natur.516..219B}. At advanced
stages of the simulation, the dominant reaction is
\begin{equation} 
  {\rm p}\htwod + {\rm o}\htwo
  \Harpoons^{k_1^+}_{k_1^-} {\rm o}\htwod + {\rm p}\htwo \; .
\end{equation}
In this case the mentioned formula can be simplified to o/p-$\htwod
\sim k_1^+/k_1^-$\,o/p-$\htwo$. 

The temporal evolution patterns of p/o-$\dtwoh$ and o/m-$\dthree$ resemble each
other, but show no obvious correlation with o/p-$\htwo$.  The lowest
energy spin modifications, meta-$\dthree$ and ortho-$\dtwoh$ are
coupled by the reaction ${\rm o}\dtwoh + {\rm HD} \leftrightarrow {\rm
  m}\dthree + {\rm o/p}\htwo$. The forward reaction is the principal
pathway to meta-$\dthree$, and the reverse reaction can only decompose
to ${\rm o}\dtwoh + {\rm HD}$ because of symmetry rules. Besides the 
fractionation reactions, the spin states of $\dtwoh$ and $\dthree$ are 
influenced by spin conversion reactions with HD: 
${\rm p}\dtwoh + {\rm HD} \rightarrow {\rm o}\dtwoh + {\rm HD}$, 
${\rm o}\dthree + {\rm HD} \rightarrow {\rm m}\dthree + {\rm HD}$
(\citealt{2004A&A...427..887F};
\citealt{2006A&A...449..621F}; \citealt{2009JChPh.130p4302H};
\citealt{2010A&A...509A..98S}).

\subsection{Simulated spectra} \label{subsec:simu_spectra}

We calculated the spectral line profiles of the ortho and para
modifications of $\htwod$ and $\dtwoh$ at different times, using the
cloud models described in Section~\ref{subsec:physical}. The evolution
of chemical abundances in the model clouds were calculated using
different initial gas compositions, determined by the duration of the
preceding ``dark cloud'' stage.  The following durations of the
initial dark cloud stage were tested: $1\times10^5$, $3\times10^5$, $5\times10^5$,
$7\times10^5$, and $1\times10^6$ yr. The dense core plus ambient cloud
evolution was followed up to $3\times10^6$ yr. The cloud models differ by
the ambient cloud parameters, and in terms of how the transition from
the dense core to the ambient cloud occurs (see
Figure~\ref{fig:crimier_ext_model}).  The purpose of the calculations
was to find a physical model that at a certain stage could reproduce
the observed $\htwod$ and $\dtwoh$ spectra.  The appropriateness of
the model and the timing was controlled with the help of diagrams
showing the time-dependence of the integrated emission/absorption line
intensity. The method is explained in Appendix A.

Typical of almost all simulations is that the ortho-$\dtwoh$ and
para-$\htwod$ absorption features are weak in the beginning but grow
stronger with time.  This is opposite to the behavior of the
ortho-$\htwod$ emission line which generally diminishes in intensity
(but in most cases remains overly strong compared with the observed
line, see below).  However, the ortho-$\dtwoh$ absorption line weakens
again and the ortho-$\htwod$ emission line typically gets stronger
toward the end of the simulation. The weakening of the ortho-$\dtwoh$
absorption is caused by the depletion of HD in the outer envelope. The
opposite tendency of ortho-$\htwod$ emission is related to a
redistribution of the ortho-$\htwod$ abundance discussed below. The
para-$\dtwoh$ emission/absorption features are always weak compared to
the noise level of the observed spectrum.

As discussed in Section~\ref{subsec:abundances}, the $\dtwoh$ abundance
grows larger when the dense core stage starts from pre-processed gas
(Figure~\ref{fig:abu_time}). The best overall agreement between
simulations and the observations is generally found when the initial
dark cloud stage has lasted $5\times10^5$ yr, and the duration of the
subsequent dense core stage is another $5\times10^5$ yr.  This time corresponds
to a minimum in the ortho-$\htwod$ emission line intensity and a
maximum in the ortho-$\dtwoh$ absorption. Lengthening the initial
stage increases the minimum intensity of the ortho-$\htwod$ line. On
the other hand, using initially atomic abundances or initial
processing times of $1-3\times10^5$ yr results in an overly weak
ortho-$\dtwoh$ absorption feature at any time of the simulation. While
the para-$\htwod$ absorption continually deepens until the end of the
simulation, the ortho-$\dtwoh$ absorption weakens at very late times.
In the simulation continuing from an initial stage of $5\times10^5$ yr,
this weakening becomes significant at $t\sim 2-3\times10^6$ yr.

The strenghtening of ortho-$\htwod$ is caused by an increase in the
ortho-$\htwod$ abundance deeper in the envelope at radii $R\sim
1500-2500$ AU where the temperature is $T \sim 18-22$ K and the
density is $n(\htwo)\sim 10^6$ cm$^{-3}$.  The effect is seen at late
times $t \gtrsim 10^6$ yr, and its occurrence depends on the initial
gas composition.  The creeping of ortho-$\htwod$ into the inner parts
of the envelope is favoured when $\hthree$ abundance is high in the
beginning of the dense core simulation, meaning an initial stage
duration of at least $3\times10^5$ yr (see also Figure~\ref{fig:a_vs_time}
in the Appendix).

\begin{figure}[]
\figurenum{8}
\plotone{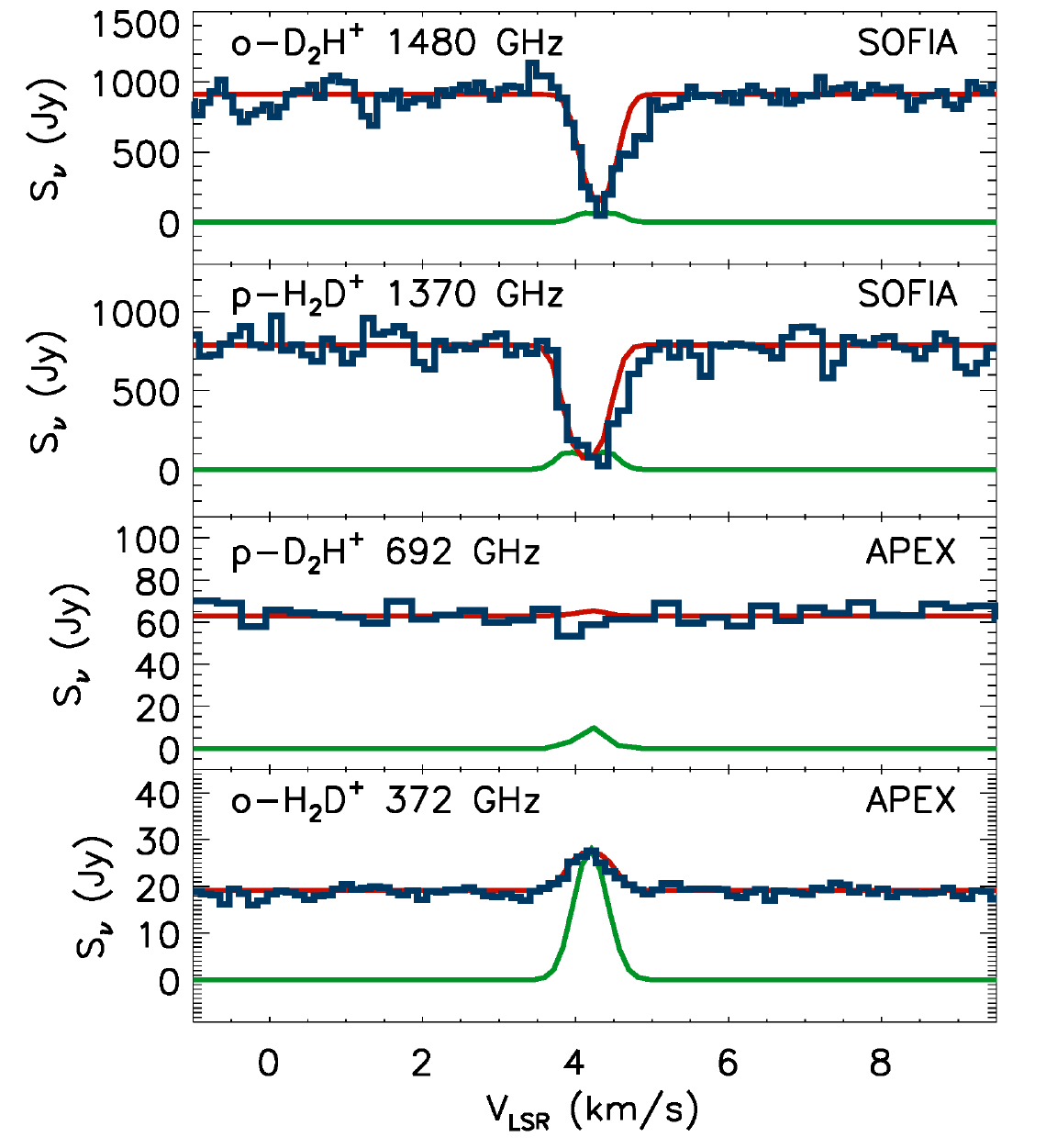}
\caption{Observed ortho-$\dtwoh$, para-$\htwod$, ortho-$\htwod$,
  and para-$\dtwoh$ spectra (histograms) toward IRAS16293.  The red
  solid curves are predictions from our best fit model when the dense
  core is assumed to be embedded in a homogeneous dark cloud with
  $n(\htwo)=10^4$ cm$^{-3}$, $T=10$ K, and there is no velocity shift
  between the foreground component and the core. The green curves show
  the predicted line emission without the continuum of the embedded
  far-IR source.}
\label{fig:model_spectra_1}
\end{figure}

The ortho-$\dtwoh$, para-$\htwod$, ortho-$\htwod$, and para-$\dtwoh$
spectra calculated from the cloud model of \cite{2014Natur.516..219B}
are shown in Figure~\ref{fig:model_spectra_1}, together with the
observed spectra.  The spectra are presented on the flux density
scale. The physical model consists of the Crimier et al. core model
surrounded by a homogeneous dark cloud with $n(\htwo)=10^4$ cm$^{-3}$,
$T=10$ K.  No velocity shift between the core and the ambient cloud is
assumed.  The model gives rather good predictions for the height of
the ortho-$\htwod$ peak and the depths of the valleys in
ortho-$\dtwoh$ and para-$\htwod$ spectra. On the other hand, it cannot
reproduce the integrated intensities and the asymmetric shapes of the
absorption spectra. 

There is a hint of blue-shifted emission in the observed
ortho-$\dtwoh$ spectrum. In view of several artefacts on this side of
the spectrum, we cannot be sure that this feature originates in the
target source. Blue-shifted emission should be attributed to infalling
gas on the far-side of the envelope.  According to the present model,
ortho-$\dtwoh$ does not show such a feature because of a very low abundance 
within the infall-radius ($R \lesssim 1000$ AU). 

More pronounced are the red-shifted shoulders observed in both
ortho-$\dtwoh$ and para-$\htwod$ absorption line profiles. These can
best be reproduced by assuming that the near side of the ambient cloud
falls toward the core at a speed of $0.2\, \kms$. We still assume that
the ambient cloud is homogeneous, but vary the temperature in the
range $10-14$ K and the density in the range $1-4 \times 10^4$ cm$^{-3}$.

\begin{figure}[]
\figurenum{9}
\plotone{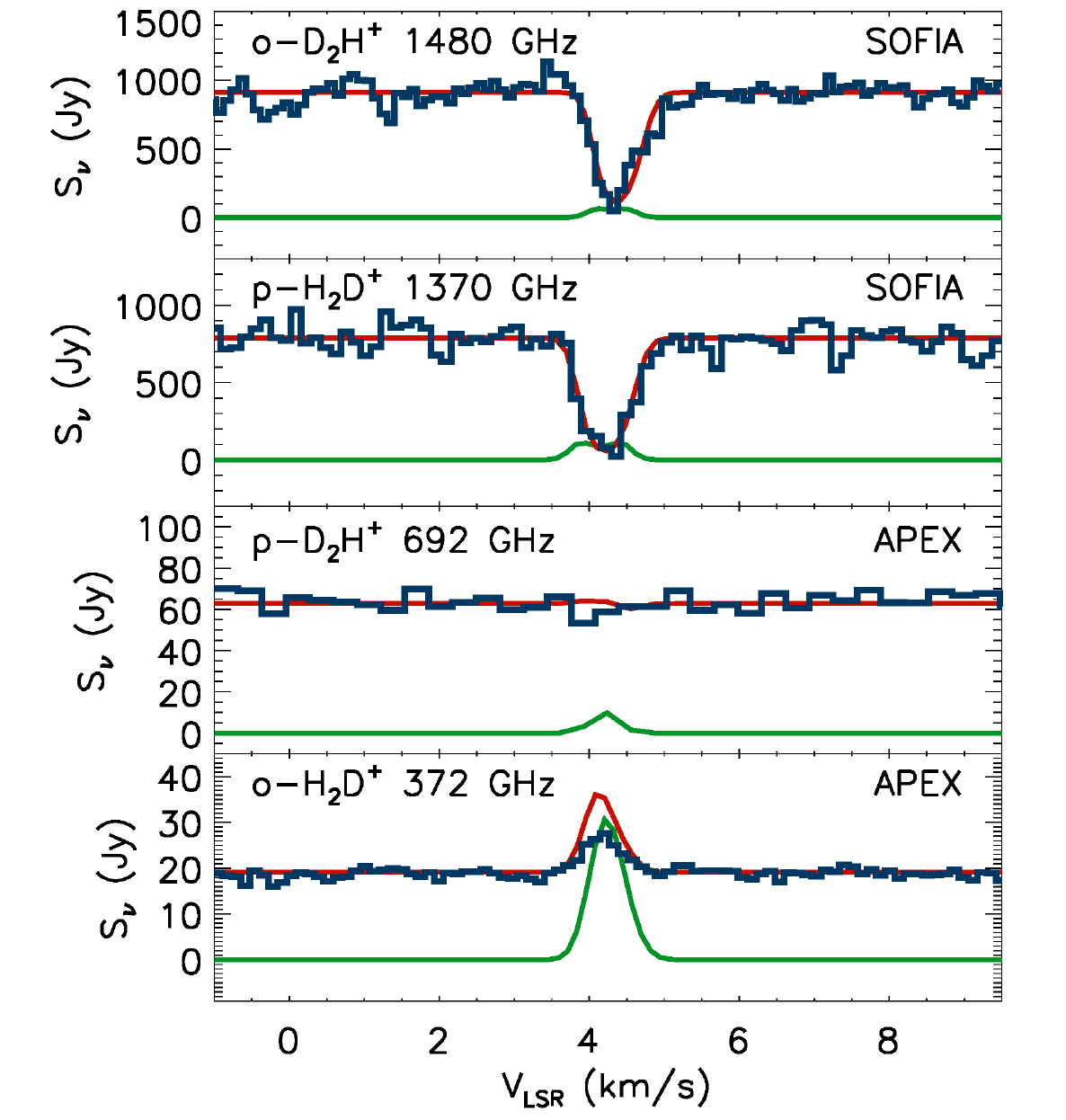}
\caption{Same as Figure~\ref{fig:model_spectra_1}, but the predicted
  spectra (red curves) are calculated using a model where the ambient
  cloud density is higher, $n(\htwo)=3\times10^4$ cm$^{-3}$, and where its
  nearside moves toward the core at a speed of $0.2\,\kms$. }
\label{fig:model_spectra_2}
\end{figure}

It turns out that the ambient cloud needs to be a little denser than
in the \cite{2014Natur.516..219B} model in order to obtain noticeable
red-shifted absorption in the ortho-$\dtwoh$ line.  On the other hand,
the intensity of the ortho-$\htwod$ line increases strongly above
$n(\htwo)\sim 3\times10^4$ cm$^{-3}$. The temperature needs to be well
below 14 K in order to cause sufficient absorption in the
ortho-$\htwod$ and ortho-$\dtwoh$ lines. The ambient cloud parameters
of the best-fit model are $T=10$ K, $n(\htwo)=3\times10^4$ cm$^{-3}$. The
predicted spectra at the time $t=5\times10^5+5\times10^5$ yr are shown in
Figure~\ref{fig:model_spectra_2}.

While the agreement between the observed and modeled absorption lines of
ortho-$\dtwoh$ and para-$\htwod$ has improved with respect to the
model of \cite{2014Natur.516..219B}, the model used for
Figure~\ref{fig:model_spectra_2} overpredicts the intensity of the
ortho-$\htwod$ emission line by a factor of $\sim 2$. Because of the velocity
shift, the foreground cloud absorbs less ortho-$\htwod$ emission from
the core as compared with the completely static model.

The situation with ortho-$\htwod$ becomes even worse when using the
extended \cite{2010A&A...519A..65C} core model, where the density and
temperature distributions are continuous. The best-fit spectra are
shown in Appendix B. In this model, there is less low-density cold gas
serving as an absorbing layer as compared with the homogeneous ambient
cloud model.  On the other hand, the model can quite well reproduce
the two absorption lines.

\subsection{Adjusting the ortho-para conversion of $\htwod$}
\label{ss:opconv}

Inspection of Figure~\ref{fig:op_rad} reveals that o/p-$\htwod$ lies a
constant factor of $\sim 2$ above the thermal value when o/p-$\htwo$
is thermalized. This is contrary to intuition and pointed us to the
fact that the rate coefficients used in the simulation lead to an
overestimate of o/p-$\htwod$. The possible overestimate could
contribute to the overly bright ortho-$\htwod$ line produced by the
present simulations. A similar tendency of over-predicting
ortho-$\htwod$ emission with the same chemistry model was observed in
a study of the starless core H-MM1 \citep{2017A&A...600A..61H}.

For reaction (1) and other reactions between the isotopologs of
$\hthree$ and $\htwo$, we have adopted the {\sl ground
  state}-to-species rate coefficients given in Table VIII of
\citet{2009JChPh.130p4302H}. The $\alpha$ and $\beta$ parameters in
the Arrhenius-Kooij expressions for the rate coefficients presented in
this table were fitted in the temperature range $5-20$
K. \citet{2013A&A...551A..38P} have extended the fits to the range
$5-50$ K. The anomalous o/p-$\htwod$ ratio is seen also at
temperatures for which the original fit is valid.

In the reaction set used, the ratio $k_1^+/k_1^-$ of the forward and
backward reactions (1) is $\sim 2.25\,\,\exp(87.7/T)$, whereas according
to statistical mechanics, the
equilibrium constant of this reaction, 
\begin{equation}
K_1(T) \, \equiv \, \frac{k_1^+}{k_1^-} \, = \, 
\frac{Q_{{\rm o}\htwod}(T)}{Q_{{\rm p}\htwod}(T)} \,
\frac{Q_{{\rm p}\htwo}(T)}{Q_{{\rm o}\htwo}(T)}
\; , 
\label{eq:K}
\end{equation}
can be approximated by $K_1(T) \sim \exp(84.1/T)$ at low temperatures
(see, e.g., Eq.~61 in \citealt{2009JChPh.130p4302H}). The use of
ground state-to-species coefficients omits in particular ``backward''
reactions involving ortho-$\htwod$ in the first excited rotational
level, $1_{10}$, which lies approximately 18 K above the ground level
($1_{11}$). The $1_{10}-1_{11}$ transition gives rise to the 372 GHz emission 
line observed here. The critical density of this transition, 
$\sim 1.3\times10^5$ cm$^{-3}$ at 15 K, is exceeded in the core model
except for the outermost shells, so it seems appropriate to include
the contribution of the $1_{10}$ level to the reaction rate. This
inclusion brings the backward rate coefficient, $k_-$, close to the
thermal value. Because reaction (1) mainly determines o/p-$\htwod$,
the factor 2.25 in the ratio $k_1^+/k_1^-$ in the adopted reaction set
explains the suprathermal o/p-$\htwod$ in the inner part of the
envelope ($R< 2500$ AU).  The situation is likely to be different for
the ``forward'' reaction. The first excited level of para-$\htwod$,
lying $\sim 66$ K above the ground state, is not easily excited in the
cool outer layers of the cloud. The critical density of the
$1_{01}-0_{00}$ transition is $\sim 8.1\times10^6$ cm$^{-3}$.

We tested the effect of increasing the rate coefficient of the
backward reaction (1) so that $k_+/k_-$ obtains the value given by
Eq.~\ref{eq:K}. As expected, the ortho-$\htwod$ abundance decreases,
and the o/p-$\htwod$ ratio follows closely the thermal ratio in the
inner parts of the core after this change.

\subsection{Applying the modified rate coefficient}

\begin{figure}[]
\figurenum{10}
\plotone{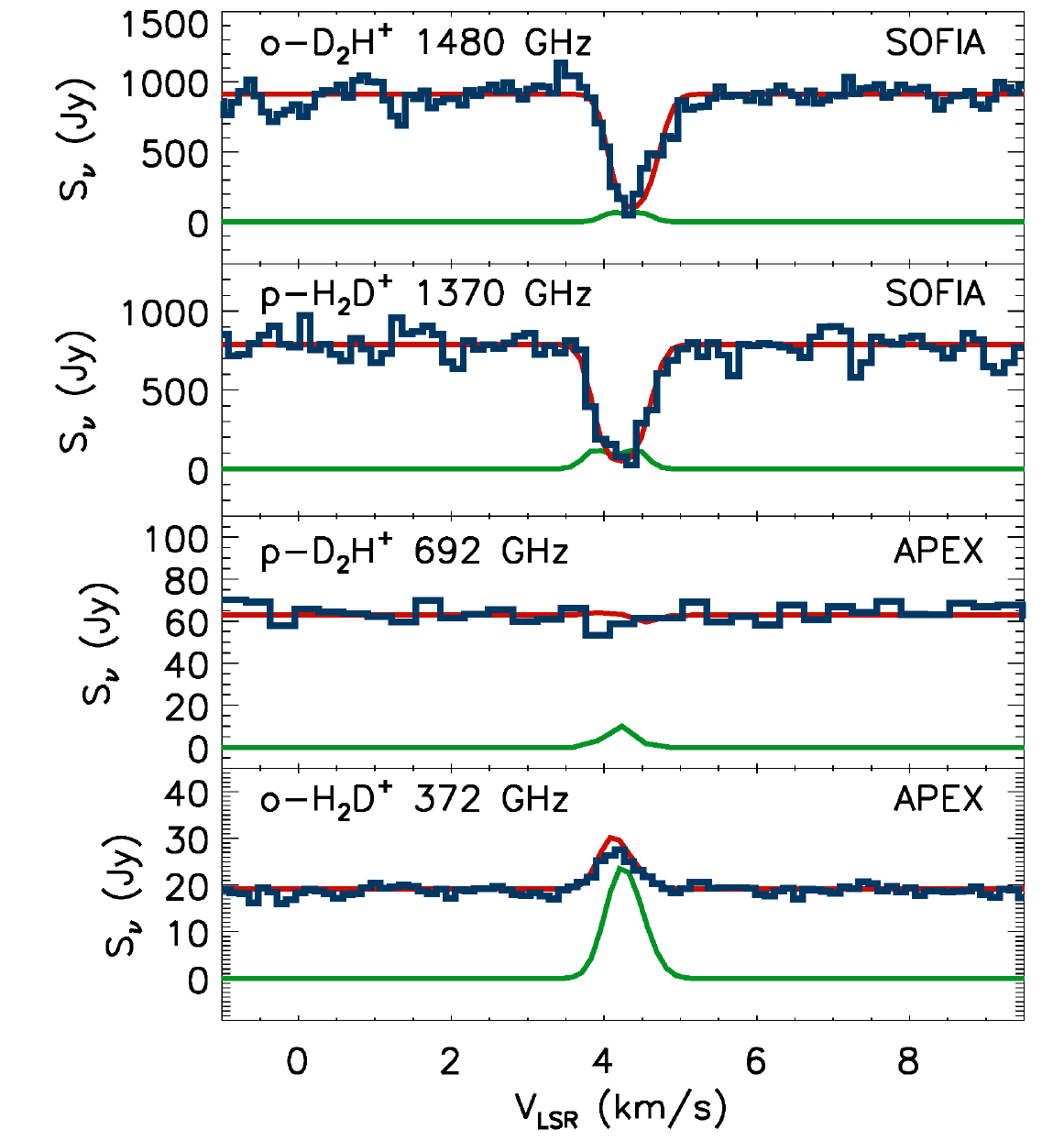}
\caption{Same as Fig.~\ref{fig:model_spectra_2}, but the predicted
  spectra (red curves) are calculated using abundances resulting from
  the modified rate coefficient for reaction (1) in the core.}
\label{fig:model_spectra_3}
\end{figure}

Spectra predicted using abundances from the modified chemistry model
in the core (but keeping to the original ground state-to-species
coefficients in the low-density ambient cloud) are shown in
Figure~\ref{fig:model_spectra_3}.  This figure presents the best-fit
spectra from the homogeneous ambient cloud with $n(\htwo)
=3\times10^4$ cm$^{-3}$ and $T=10$ K. The simulation time is again
$5\times10^5 + 5\times10^5$ yr. The corresponding model spectra from
the extended Crimier et al. core model are shown in Appendix B. For
both physical models the agreement with the observed ortho-$\htwod$
line has improved, but the homogeneous ambient cloud model gives a
clearly better fit.

Reaction (1) is probably one of the most important cases in which
ground state-to-species coefficients can deviate from the
species-to-species coefficients, when the gas density is high
enough. Figure~\ref{fig:op_rad} suggests furthermore that this effect
needs to be taken into account in reactions determining the
o/m-$\dthree$ ratio. In the present paper, we do not follow up this
matter, but intend to do so in a future study (Sipil\"a et al., in
preparation).

We conclude this section by listing in Table~\ref{tab:abundances} the
average fractional abundances of the observed molecules according to
the models used to produce the spectra shown in
Figs.~\ref{fig:model_spectra_3}. We give separately the average
abundances in the outer envelope ($R=2500-6900$ AU) and in the
homogenous foreground dark cloud (extending to $R\sim 20000$ AU).  For
the extended Crimier et al. power-law model, the abundances are
averages over the range $R=2500-10000$ AU.
The abundances in the core ($R \leq 6900$ AU) are calculated using the
modified $k_-$ coefficient for reaction (1).  Also given in this
table, in columns 4, 5, and 7, are the fractional abundances and their
uncertainties derived using the observed spectra and the physical
models assuming that the abundances are constant in the regions where
the observed emission/absorption is likely to originate, and zero
elsewhere (that is, within the radial distance $R=2500$ AU from the
center).

According to the Crimier et al. core plus ambient dark cloud model,
the average fractional ortho-$\dtwoh$ and para-$\htwod$ abundances are
higher in the outer envelope than in the ambient cloud. Column
densities give, however, a better idea of which component dominates
the absorption features. The ratios of the beam-averaged column
densities are $\sim 4$ for ortho-$\dtwoh$ and $\sim 6$ for
para-$\htwod$ in favor of the core according to this model.  As
suggested by the simulated spectra shown in
Figure~\ref{fig:model_spectra_3}, the situation is opposite for
ortho-$\htwod$: the beam-averaged column density in the ambient cloud
is three times higher than that in the core.

\setcounter{table}{1}
\begin{table*}[hbtp] 
\renewcommand{\thetable}{\arabic{table}}
\centering
\caption{Fractional abundances according to the models producing 
  the spectra shown in Figure~\ref{fig:model_spectra_3}. 
  For the homogeneous ambient cloud model, abundances in the outer envelope 
  $R=2500-6900$ AU and the ambient dark cloud are given separately. 
  For the extended Crimier et al. model, the abundances are averages 
  over the range $R=2500-10000$ AU. Columns 4, 5, and 7 give abundances derived 
  from the observed lines using the two 
  physical models and assuming constant fractional abundances in the cloud 
  components where the observed signal is likely to arise from.} 
\begin{tabular}{lllllll}
\tablewidth{0pt}
\hline
\hline
Species & 
\multicolumn{4}{c}{Crimier + ambient dark cloud} & \multicolumn{2}{c}{Extended Crimier} \\
& \multicolumn{2}{c}{Chemistry model} &
\multicolumn{2}{c}{Constant X}     & Chemistry model & Constant X  \\
  & Envelope & Ambient & Envelope & Ambient  & & \\
\multicolumn{1}{c}{(1)} & \multicolumn{1}{c}{(2)} & \multicolumn{1}{c}{(3)} &
\multicolumn{1}{c}{(4)} & \multicolumn{1}{c}{(5)} & \multicolumn{1}{c}{(6)} & 
\multicolumn{1}{c}{(7)} \\ \hline
ortho-$\dtwoh$&   $1.6\times10^{-9}$&$1.1\times10^{-9}$ &$(1.4\pm 0.4)\times10^{-9}$& $(1.0\pm0.3)\times10^{-9}$  &  $1.5\times10^{-9}$ &$(1.7\pm0.3)\times10^{-9}$    \\
para-$\dtwoh$ &   $1.8\times10^{-10}$&$1.9\times10^{-10}$&$<4\times10^{10}$          & $<4\times10^{10}$          &  $2.1\times10^{-10}$&$<4\times10^{-10}$             \\
para-$\htwod$ &  $1.6\times10^{-9}$ &$8.6\times10^{-10}$&$(1.9\pm0.7)\times10^{-9}$ & $(1.0\pm0.4)\times10^{-9}$  &  $1.7\times10^{-9}$ &$(1.6\pm0.4)\times10^{-9}$     \\
ortho-$\htwod$&  $1.1\times10^{-10}$&$1.0\times10^{-9}$ &$(1.0\pm0.1)\times10^{-10}$& $(1.3\pm0.4)\times10^{-9}$  &  $1.6\times10^{-10}$&$(7.5\pm0.6)\times10^{-11}$    \\
\hline
\end{tabular}
\label{tab:abundances}
\end{table*}

\section{Discussion and conclusions}   \label{sec:discussion}

The deepest absorption features of ortho-$\dtwoh$ and para-$\htwod$
toward IRAS16293 are found at a LSR velocity of $v_{\rm LSR} = 4.2\,
\kms$.  This velocity agrees with the velocities of the previously
observed absorption features in the spectra of D$_2$CO
\citep{2000A&A...359.1169L}, HDO (\citealt{2004ApJ...608..341S};
\citealt{2012A&A...539A.132C}), and D$_2$O
(\citealt{2010A&A...521L..31V}; \citealt{2013A&A...553A..75C}).  In
these previous studies the most prominent absorption has been
attributed to the cold outer envelope of the protostellar core. This
idea is supported by the present modeling results which indicate that
most of ortho-$\dtwoh$ and para-$\htwod$ absorption occurs in the
relatively dense outer envelope. The ortho-$\dtwoh$ and para-$\htwod$
spectra are very different from the spectra of deuterated water, for example,
which show toward the same position broad emission from the in-falling
warm inner core, in addition to the absorption dip at $v_{\rm LSR}
\sim 4.2\, \kms$ \citep{2012A&A...539A.132C}.

The origin of the ortho-$\htwod$ emission is less clear because this
line is likely to be substantially self-absorbed in the ambient
cloud. The contribution of the ambient or ``foreground'' cloud
\citep{2013A&A...553A..75C} can be partially separated thanks to a
line-of-sight velocity gradient which gives rise to asymmetric
ortho-$\dtwoh$ and para-$\htwod$ absorption profiles.  The velocity
shift between the foreground cloud and the core was not taken into
account in \citet{2014Natur.516..219B}. This made the modeled
ortho-$\htwod$ emission lines generally weaker than in the present
calculations. To serve as an absorbing component, the foreground cloud
must have a moderate density of the order of $10^4$ cm$^{-3}$, and to
develop a high ortho-$\htwod$ abundance of $\sim 10^{-9}$ at this
density, the cloud must be cold. Our spectral line simulations give
the best agreement with the observed lines when the kinetic
temperature of the foreground cloud is $T \sim 10$ K and the density
is $n(\htwo) \sim 3\times10^4$ cm$^{-3}$. These numbers are not far from
the estimates \cite{1987A&A...177L..57M} obtained for the temperature
and the density of the ambient cloud using the inversion lines of
$\ammo$: $T=11.5^{+3}_{-1}$ K, $n(\htwo) \geq 2-3\times10^4$ cm$^{-3}$.

The ortho-$\dtwoh$ abundance in the cool outer envelope of IRAS16293
is similar to that derived for para-$\htwod$. As these two species are
found to be dominant forms over the variants para-$\dtwoh$ and
ortho-$\htwod$, respectively, the conclusion is that the total
abundances of $\dtwoh$ and $\htwod$ are approximately equal. This
agrees with what is expected for deuterium chemistry in cool,
CO-depleted gas since the pioneering modeling works of
\citet{2003ApJ...591L..41R} and \citet{2004A&A...418.1035W}. The
mentioned studies predict furthermore that in these conditions also
$\dthree$ should be very abundant, perhaps the most abundant of the
$\hthree$ isotopologues.  This is because all three reactions in the
sequence $\hthree + {\rm HD} \rightarrow \htwod + \htwo$, $\htwod +
{\rm HD} \rightarrow \dtwoh +\htwo$, $\dtwoh + {\rm HD} \rightarrow
\dthree + \htwo$ are exothermic by about 200 K and should proceed
rapidly in cold gas.  In the present model, $\dthree$ becomes the
most abundant ion in the coolest parts of the outer envelope ($R\sim
3000-6000$ AU), but there are no large differences between the
fractional abundances of $\hthree$, $\htwod$, $\dtwoh$, $\dthree$ and
H$^+$ there.

The symmetric ions $\hthree$ and $\dthree$ are spectroscopically
  identifiable through their rovibrational transitions in the
  infrared. Since the first interstellar detection by
  \cite{1996Natur.384..334G}, $\hthree$ absorption lines have been
  detected both in diffuse and dense interstellar gas in front of
  strong infrared sources (e.g., \citealt{2014ApJ...786...96G}). The
  $\dthree$ ion has not yet been detected. As discussed by
  \cite{2013ChRv..113.8738O}, the difficulty is to find a cold and
  dense region with a high degree of deuterium fractionation that 
  obscures a luminous infrared source.  The other problem is
  that the fundamental rovibrational lines of $\dthree$ lie at
  wavelengths where the atmosphere is opaque.  The predicted
  maximum column density of the most abundant, meta ($A_1$) spin
  modification of $\dthree$ in the envelope of IRAS 16293 is
  approximately $10^{13}\,{\rm cm}^{-2}$. Using Table 3 of
  \cite{2004A&A...427..887F}, based on spectroscopic data presented by
  \cite{2004MNRAS.354..161R}, one finds that the quoted column density
  corresponds to a peak optical thickness of $\tau=0.5$ of the
  ground-state absorption line of meta-$\dthree$ at 5.3 $\mu$m micron.
  The assumption here is that the line width is 1 $\kms$. (The meta and ortho
  appellations of $\dthree$ in \cite{2004A&A...427..887F} are
  interchanged with respect to the nomenclature adopted here.) Despite
  significant optical thickness, the line cannot be detected toward
  IRAS16293 because of the very weak continuum at 5.3 $\mu$m. The
  source is not detected in the IRAC bands of the Spitzer Space
  Telescope; the Band 3 (5.7 $\mu$m) surface brightness at this
  position is approximately 0.1~mJy$(\,\square\arcsec)^{-1}$.  The
  line is currently accessible to the Echelon-Cross-Echelle
  Spectrograph
  (EXES)\footnote{www.sofia.usra.edu/science/instruments/exes} onboard
  SOFIA. Despite the difficulties mentioned above, it seems
  worth while to attempt detection toward a luminous embedded Young
  Stellar Object, if there is reason to believe that the
  meta-$\dthree$ column density in the foreground gas is of the order
  of $10^{12}-10^{13}$ cm$^{-2}$ or more.

The spin ratios of the $\hthree$ isotopologs are predicted to
exhibit overall radial gradients in the envelope of IRAS16293
because of a steep gradient in the kinetic temperature. The lower
energy nuclear spin modifications, ortho-$\dtwoh$ and para-$\htwod$,
should be strongly favored in the cool outer envelope. The non-thermal
o/p-$\htwo$ ratio causes, however, that also o/p-$\htwod$ and
o/p-$\dtwoh$ deviate from their thermal equilibrium values there
(Figure~\ref{fig:op_rad}). As discussed in
\citet{2014Natur.516..219B}, and above in
Section~\ref{subsec:abundances}, o/p-$\htwod$ correlates closely to
o/p-$\htwo$, which thermalizes quickly at temperatures above 20 K (at
radii $R \leq 2000$ AU in the present source model), but hardly ever
reaches thermal equilibrium in cold gas.  The thermalization of
o/p-$\htwo$ in warm gas occurs efficiently in proton exchange
reactions with ${\rm H^+}$ (e.g., \citealt{2011ApJ...729...15C}). At
low temperatures prevailing in the outer envelope, o/p-$\htwo$
approaches slowly the thermal value through reactions with $\hthree$
and its isotopologs. This process is still slower in the ambient
cloud, which likewise is cool but has an order of magnitude lower
density. In this component, ortho-para conversion reactions are
slow compared to the replenishment of $\htwo$ from
grains. Consequently, the o/p-$\htwo$ ratio is clearly suprathermal,
and so is o/p-$\htwod$.

The o/p-$\dtwoh$ ratio depends indirectly on o/p-$\htwo$.  For
ortho-$\dtwoh$, the principal formation and destruction pathways are
the following fractionation reactions in the left-to-right direction:
\begin{eqnarray}
\htwod + {\rm HD} & \; \leftrightarrow \; & \dtwoh + \htwo \; , 
\label{eq:plusd2h+} \\
\dtwoh + {\rm HD} & \; \leftrightarrow \; & \; \dthree \; \;  + \; \htwo \; .
\label{eq:minusd2h+} 
\end{eqnarray}
The forward reactions are strongly exothermic when the products are in
their lowest energy states. Right-to-left reactions are, however,
relatively fast in the presence of ortho-$\htwo$. Consequently,
ortho-$\dtwoh$ is destroyed in the backward reaction
(\ref{eq:plusd2h+}) and, to a lesser extent, created in the reverse
reaction (\ref{eq:minusd2h+}), but significantly only if $\htwo$ is in
the ortho form. In contrast, the higher-energy para-$\dtwoh$ is
efficiently destroyed also by para-$\htwo$ through the reverse
reaction (\ref{eq:plusd2h+}). The net effect is that o/p-$\dtwoh$
shows a weak anti-correlation with o/p-$\htwo$
(Figure~\ref{fig:op_rad}).  However, no simple formula relating
o/p-$\dtwoh$ to o/p-$\htwo$ in a wide range of physical conditions can
be presented.

The spin ratios are in general predicted to approach slowly the
thermal values.  In the cool outer parts of the envelope, $R\sim
2500-6900$ AU, this tendency is clear for o/p-$\htwod$ which is driven
by o/p-$\htwo$, whereas o/p-$\dtwoh$ remains nearly constant in the
time interval $10^5 - 10^6$ yr (Figure~\ref{fig:op_time}).  This
suggests that the o/p-$\dtwoh$ ratio is not as trustworthy as a
chemical clock as o/p-$\htwod$.

The gas we are probing with $\dtwoh$ and $\htwod$ makes up the outer
envelope of the core which has only recently, probably a few times
$10^4$ years ago, formed the Class 0 protostellar system IRAS16293
A/B. While the interior parts of the core within a few 100 AU are
strongly influenced by the newly born stars
(e.g. \citealt{2016A&A...595A.117J}), the outer envelope is likely to
represent material accumulated during the pre-stellar stage.  Based on
the ortho/para-$\htwod$ abundance ratio in this component,
\cite{2014Natur.516..219B} concluded that the current star formation
activity in IRAS16293 was preceded by a period of very slow
contraction; the best agreement with the observed para- and
ortho-$\htwod$ lines was reached with a model where the chemical age
of the core was $10^6$, and this stage was preceded by $10^6$ yr of
ambient cloud evolution.

The combination of the $\dtwoh$ observations with the previous
$\htwod$ data, and the refinement of the cloud model to include the
small velocity difference between the core and the ambient cloud,
impose a change with respect to our previous age estimate.  The
ortho-$\dtwoh$ abundance in the outer envelope reaches a maximum
somewhere between $3\times10^5$ and $10^6$ yr, depending on the initial
abundances, and decreases again at very late times when the total
deuterium abundance in the gas phase starts to diminish because of
accretion onto grains. In contrast, the average ortho-$\htwod$
abundance in the core has a minimum close to the ortho-$\dtwoh$
maximum.  In our previous model presented in
\cite{2014Natur.516..219B}, the late-time brightening of
ortho-$\htwod$ line was hidden by absorption in the ambient cloud.
The brightening is caused by an increase in the ortho-$\htwod$
abundance in the luke-warm ($T\sim 20-22$ K) layers of the envelope at
radii $R\sim 1500-2500$ AU, while at the same time the ortho-$\htwod$
abundance continues to decrease in the cool outer parts.  The
late-time increase in the inner envelope does not happen to other
species considered here, and it does not happen to ortho-$\htwod$
either, if the dense core simulation is started from ``fresh'' gas,
that is, when the duration of the initial stage is less than $\sim
3\times10^5$ yr. In this case, however, the ortho-$\dtwoh$ absorption
never reaches the observed level (see Figure~\ref{fig:a_vs_time}).  We
note that the modification of the rate coefficient $k_1^-$ of reaction
(1) to take into account ortho-$\htwod$ ions in the first excited
rotational level $1_{01}$ (Section~\ref{ss:opconv}) does not affect
the timing of the ortho-$\htwod$ minimum nor the other time-scales
derived here.

The diverging temporal behaviors of ortho-$\htwod$ and ortho-$\dtwoh$,
combined with the continuous increase in the para-$\htwod$ abundance
with time, put tighter constraints on the chemical age than the
o/p-$\htwod$ ratio alone. Using the physical model of
\cite{2010A&A...519A..65C} for the IRAS16293 core, and varying the
ambient cloud conditions within a reasonable range agreeing with
previous observations, the observed $\dtwoh$ and $\htwod$ spectra can
be best reproduced when the dense core age is $\sim 5\times10^5$ yr
and the ambient cloud is twice as old, $\sim 10^6$ yr. These age
estimates based on chemical evolution in a static dense core are still
long compared to the age of the embedded protostars, and support the
general conlusion of \cite{2014Natur.516..219B} that the current star
formation activity in IRAS16293 has been preceded by a long
pre-stellar phase. The age of the core is comparable with the
  average prestellar core lifetime in Ophiuchus and in two other nearby
  clouds estimated by \cite{2008ApJ...684.1240E}.  Based on the approximately
  equal numbers of starless cores and embedded protostars, and the
  total duration of the embedded (Class 0 and I) protostar stage of
  $0.54$ Myr \citep{2009ApJS..181..321E}, they obtain an average prestellar
  lifetime of $\sim 5\times10^5$ yr. \cite{2008ApJ...684.1240E} estimate this
  value to be uncertain by a factor of 2 in either direction. The
  chemical age of IRAS16293 derived here is likely to be a lower
  limit. In a more realistic model, where the core gradually reaches
its present density, the increase in the o/p-$\dtwoh$ ratio and the
decrease in the o/p-$\htwod$ ratio would proceed more slowly than in
the model we are using, and probably also the resulting age estimate
would need to be adjusted upwards.

\vspace{2mm}

The authors thank the anonymous referee for raising questions that
helped to improve this paper.  Stratospheric Observatory for Infrared
Astronomy (SOFIA) is jointly operated by the Universities Space
Research Association, Inc. (USRA), under NASA contract NAS2-97001, and
the Deutsches SOFIA Institut (DSI) under DLR contract 50 OK 0901 to
the University of Stuttgart. GREAT is a development by the Max Planck
Institut f{\"u}r Radioastronomie (MPIfR) and the KOSMA/Universit\"at
zu K\"oln, in cooperation with the MPI f\"ur Sonnensystemforschung and
the DLR Institut f\"ur Planetenforschung.  The Atacama Pathfinder
Experiment, APEX, is a collaboration between the MPIfR, the Onsala
Space Observatory (OSO), and the European Southern Observatory (ESO).
J.H. thanks the Max-Planck Institut for Extraterrestrial Physics for
generous support. J.H. and M.J. acknowledge financial support from the
Academy of Finland grant 258769. P.C. acknowledges support from the
European Research Council (ERC; project PALs 320620). The Cologne and
Bonn groups were supported by the Deutsche Forschungsgemeinschaft
through SFB 956.

\appendix
\section{Determining the best-fit model and time}
\label{sec:timing}

\begin{figure*}[htbp]
\figurenum{11}
\unitlength=1mm
\begin{picture}(160,95)
\put(0,0){
\begin{picture}(0,0) 
\includegraphics[width=9cm]{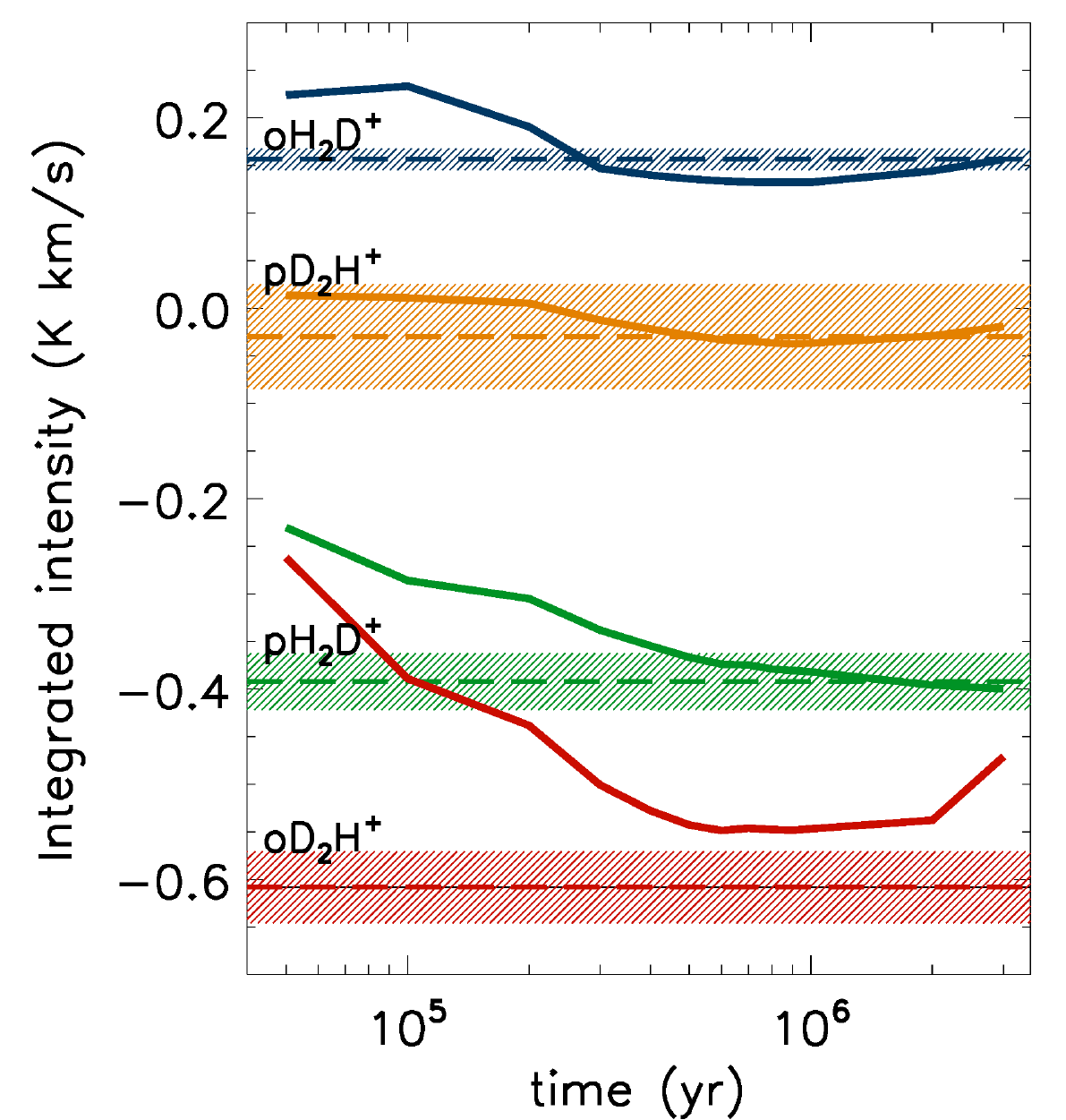}
\end{picture}}
\put(90,0){
\begin{picture}(0,0) 
\includegraphics[width=9cm]{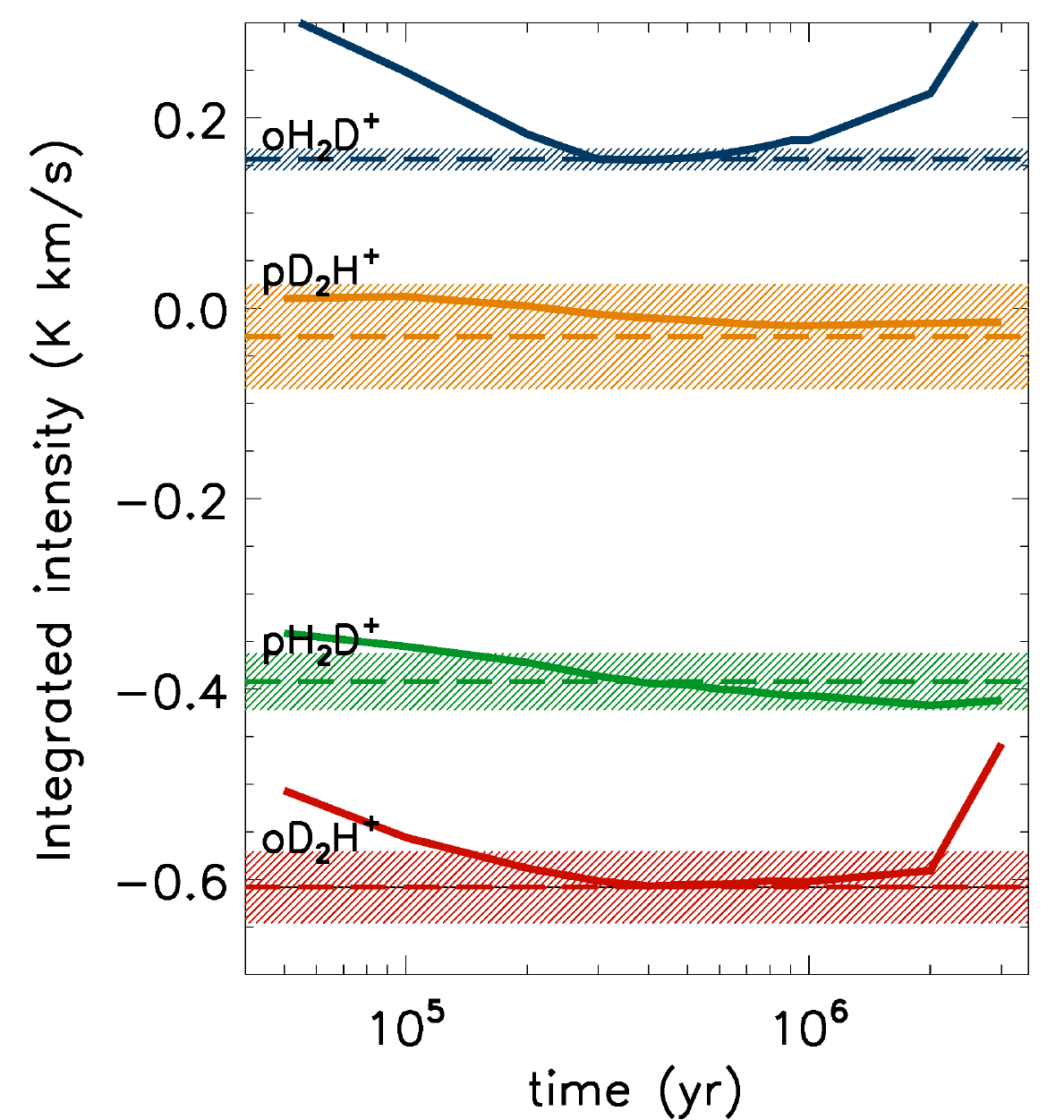}
\end{picture}}
\put(26,89){\makebox(0,0){\large \bf a)}}
\put(116,89){\makebox(0,0){\large \bf b)}}
\end{picture}
\caption{Integrated (continuum subtracted) line
  areas as functions of time after the beginning of the dense core
  phase (solid lines) for the Crimier et al. core plus ambient dark 
 cloud model. These are shown for two different initial gas 
compositions which are taken a dark cloud stage at $3\times10^5$ yr (a) 
 and $5\times10^5$ yr (b).  The hatched horizontal bars describe the
  observed integrated intensities and their 1-$\sigma$ errors. The
  ambient cloud is assumed to be homogeneous with $T=10$ K and
  $n(\htwo)=3\times10^4$ cm$^{-3}$. Both calculations are done using 
  a modified backward rate coefficient of reaction (1) explained
  in Section~\ref{ss:opconv}.}
\label{fig:a_vs_time}
\end{figure*}

We monitored the influence of the physical parameters and the initial
abundances on the predicted spectral line intensities using diagrams
such as those presented in Figure~\ref{fig:a_vs_time}, which show the
continuum-subtracted integrated line intensities of ortho-$\htwod$,
para-$\dtwoh$, para-$\htwod$, and ortho-$\dtwoh$ as functions of
time. For para-$\htwod$ and ortho-$\dtwoh$, a decreasing curve in
the diagram means deepening absorption and an increasing
abundance. The predicted spectra agree approximately with the
observations if all the four curves simultaneously pass or come close
to the center lines of the horizontal bars representing the observed
integrated intensities. In the parameter space probed here, such cases
are exclusively found when the initial stage duration is $5\times10^5$ yr,
and when the ambient cloud densities and temperatures are in narrow
ranges of $n(\htwo) = 2-3 \times10^4$ cm$^{-3}$ and $T=10-11$ K,
respectively. To illustrate the effect of the initial abundances, we
show in Figure~\ref{fig:a_vs_time} the results for simulations starting
from the times $3\times10^5$ yr (a) and $5\times10^5$ yr (b) of the dark cloud
stage. The ambient cloud density is $3\times10^4$ cm$^{-3}$ and the
temperature is 10 K. These models use the modified rate coefficient $k_1^-$ 
of reaction (1). Otherwise the ortho-$\htwod$ intensity would not reach
a sufficiently low level in simulation (b).

As suggested by the diagrams, the simulated spectra agree best with
the observations at the time $5\times10^5 + 5\times10^5$ yr. The predicted
spectra from this time are shown in 
Figure~\ref{fig:model_spectra_3}.  There is however,
considerable uncertainty in the timing. All the spectra in the time
interval $3\times10^5 - 7\times10^5$ yr, corresponding the time when all the
curves of Figure~\ref{fig:a_vs_time} are inside the hatched areas, agree
reasonably well with the observations.

\section{Predicted spectra for continuous density 
and temperature
  distributions}

\begin{figure*}[htbp]
\figurenum{12}
\gridline{\fig{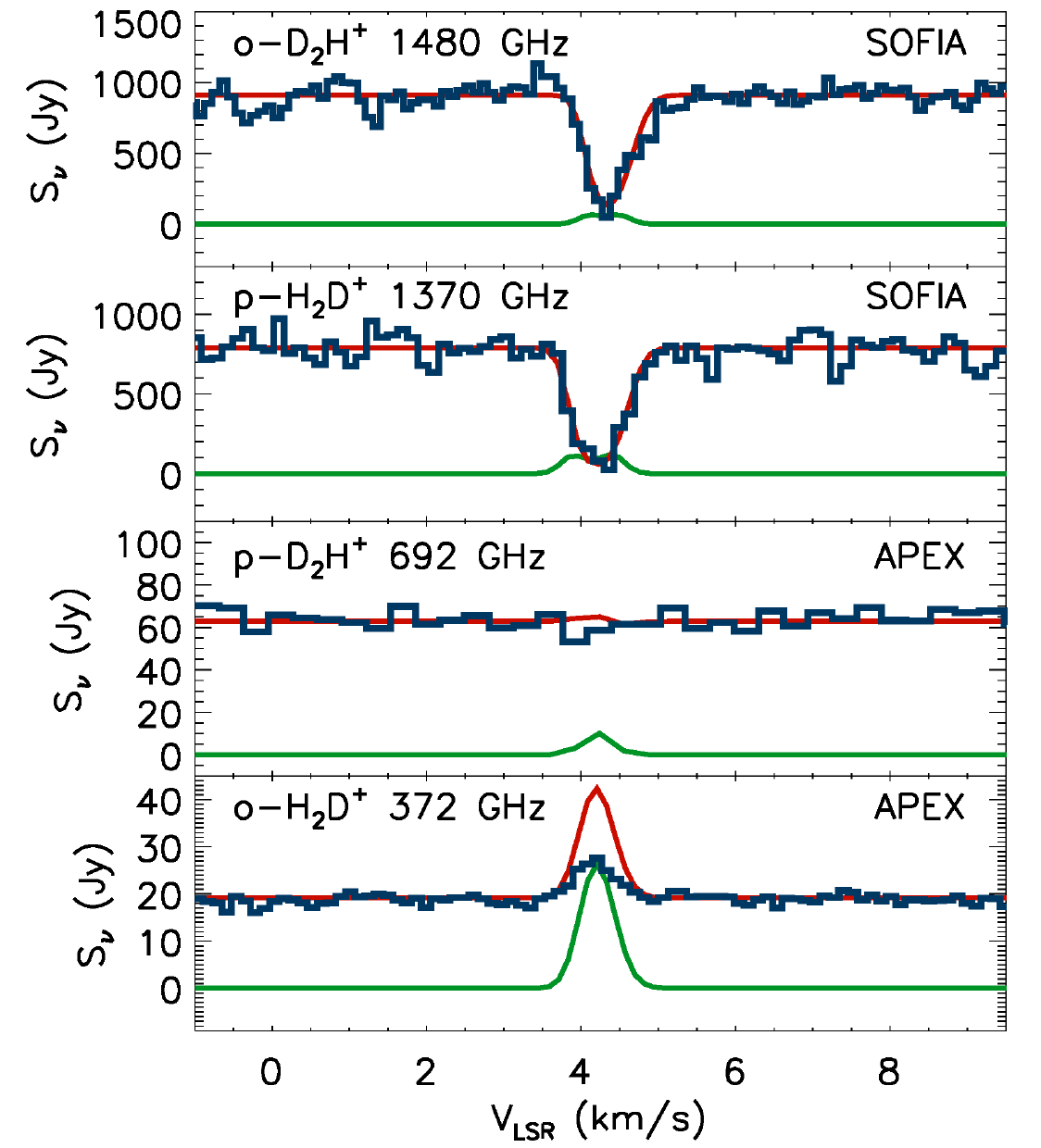}{0.48\textwidth}{(a)}
          \fig{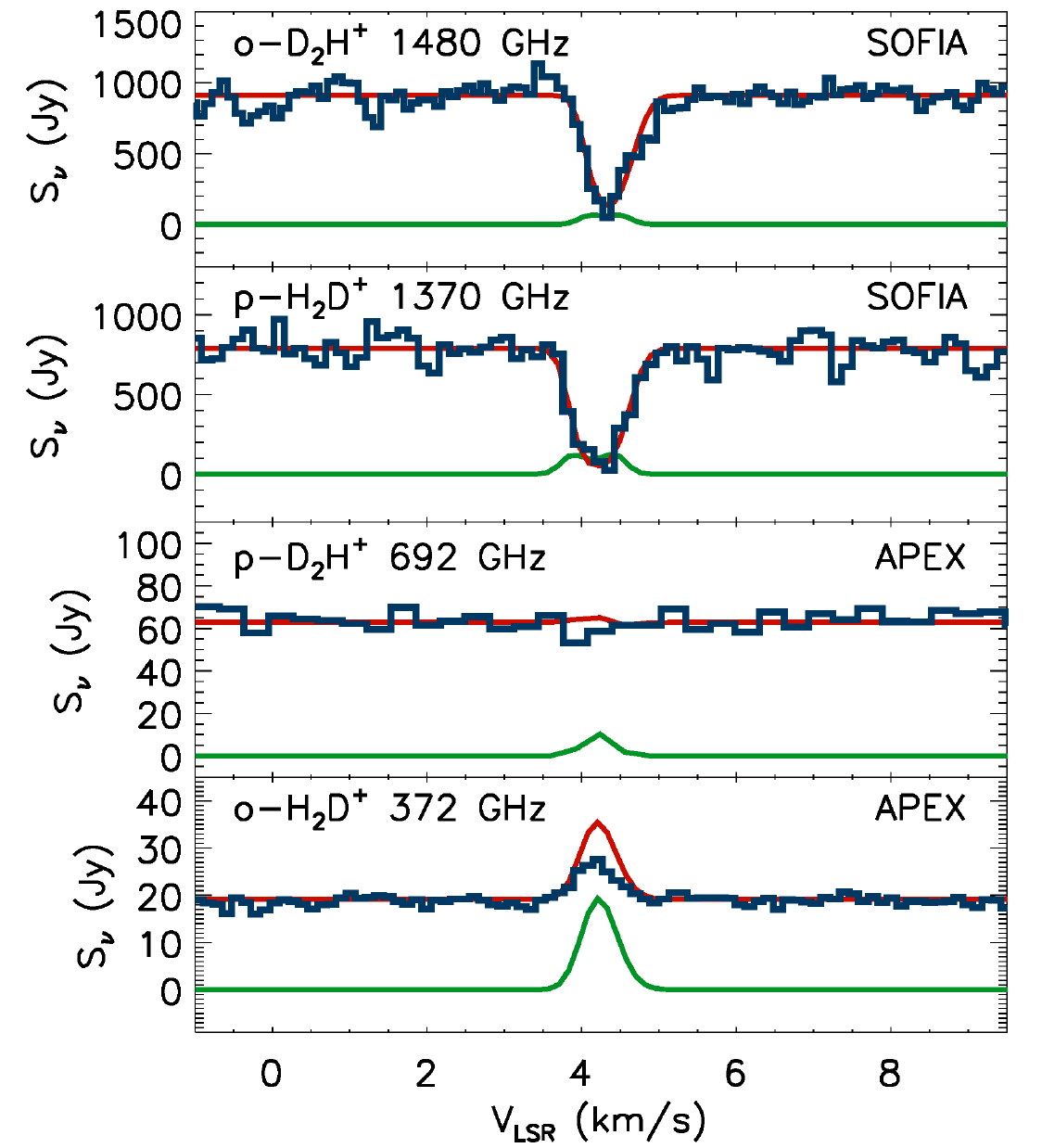}{0.48\textwidth}{(b)}}
        \caption{Observed and modeled ortho-$\dtwoh$, para-$\htwod$,
          ortho-$\htwod$, and para-$\dtwoh$ spectra toward IRAS16293
          when the cloud density structure is assumed to follow a
          power-law ($n(\htwo)\propto r^{-1.8}$) up to the
          surface. The best-fit model spectra on the left (a, red
          solid curves) are from calculations using the original rate
          coefficients in the chemistry model. Those on the right (b)
          correspond to the modified rate coefficient $k_1^-$ of
          reaction (1). The modification affects mainly the
          ortho-$\htwod$ abundance and the intensity of the 372 GHz
          line.}
\label{fig:model_spectra_ext}
\end{figure*}

The best-fit spectra from the extended core model of
\cite{2010A&A...519A..65C} are shown
Figure~\ref{fig:model_spectra_ext}. The density and temperature
distributions of the ambient cloud are shown with dashed lines in
Figure~\ref{fig:crimier_ext_model}. The near-side of the ambient cloud
is assumed to move at a velocity of $+0.2\,\kms$ with respect to the
core. The simulation time is $t=5\times10^5 + 5\times10^5$ yr (dark cloud +
dense core). The spectra corresponding to the abundances calculated
using the original ground state-to-species rate coefficients are shown
on the left (a) and those using the modified backward rate of reaction
(1) are shown on the right (b).

\bibliographystyle{aasjournal} 

\bibliography{harju}


\end{document}